# Complex vectorial optics through gradient index lens cascades

Chao He[1,*], Jintao Chang[2,3], Qi Hu[1], Jingyu Wang[1], Jacopo Antonello[1], Honghui He[3], Shaoxiong Liu[4], Jianyu Lin[5], Ben Dai[6], Daniel S. Elson[5], Peng Xi[7], Hui Ma[2,3] and Martin J. Booth[1,*]

[1]*Department of Engineering Science, University of Oxford, Parks Road, Oxford, OX1 3PJ, UK*

[2]*Department of Physics, Tsinghua University, Beijing 100084, China*

[3]*Shenzhen Key Laboratory for Minimal Invasive Medical Technologies, Institute of Optical Imaging and Sensing, Graduate School at Shenzhen, Tsinghua University, Shenzhen 518055, China*

[4]*Shenzhen Sixth People's Hospital (Nanshan Hospital) Huazhong University of Science and Technology Union Shenzhen Hospital, Shenzhen 518052, China*

[5]*Hamlyn Centre for Robotic Surgery, Institute of Global Health Innovation, Imperial College London, London SW7 2AZ, UK*

[6]*School of Data Science, City University of Hong Kong, Kowloon, Hong Kong, China*

[7]*Department of Biomedical Engineering, College of Engineering, Peking University, Beijing 100871, China*

[*]*Corresponding authors: chao.he@eng.ox.ac.uk; martin.booth@eng.ox.ac.uk*

**Graded index (GRIN) lenses are commonly used for compact imaging systems. It is not widely appreciated that the ion-exchange process that creates the rotationally symmetric GRIN lens index profile also causes a symmetric birefringence variation. This property is usually considered a nuisance, such that manufacturing processes are optimized to keep it to a minimum. Here, rather than avoiding this birefringence, we understand and harness it by using GRIN lenses in cascade with other optical components to enable extra functionality in commonplace GRIN lens systems. We show how birefringence in the GRIN cascades can generate vector vortex beams and foci, and how it can be used advantageously to improve axial resolution. Through using the birefringence for analysis, we show that the GRIN cascades form the basis of a new single-shot Müller matrix polarimeter with potential for endoscopic label-free cancer diagnostics. The versatility of these cascades opens up new technological directions.**

## Introduction

Graded index (GRIN) lenses have a varying refractive index profile that enables focusing or imaging through a compact rod-like structure[1-8]. The lower mass and size of GRIN lenses mean that are preferable to other conventional optics in many applications: shorter ones can be used as coupling lenses in fibre systems or in waveguide-based on-chip devices[1, 2, 6]; longer counterparts are normally treated as relay lenses or objective lenses that can form biopsy needles for disease diagnosis. They can also be found in other compact systems, such as fluorescence micro-endoscopes for application in deep tissue imaging[1, 4, 8].

Procedures for fabrication of GRIN lenses are well established: an ion-exchange process creates a rotationally symmetric index profile in the glass rod. There is however an undesirable side-effect: the process also introduces a concomitant intrinsic birefringence that maintains the same rotational symmetry[9, 10] (see Figs. 1a to 1c). This gradually changing birefringence profile exhibits the following properties: 1) the magnitude of the retardance is constant at a given radius, 2) the retardance increases with increasing radius; and 3) the slow axis is oriented in the radial direction. These properties mean that the GRIN lens behaves like a spatially variant waveplate array, providing a continuum of birefringence states that can manipulate the polarization and phase of a light beam (see Fig. 1d and details in **Supplementary Note 1**).

Here, through better understanding of these phenomena, we have drawn upon these previously undesirable GRIN lens properties to build new light manipulation structures to extend the applications of traditional GRIN lens systems (GLS). We call these structures “GRIN lens cascades”. As illustrated in Fig. 1e, these cascades comprise one or more GRIN lenses with interstitial optical elements, including polarizers and waveplates. Such combinations of elements introduce myriad possibilities for structuring the phase and polarization profiles of beams and foci. We describe here several aspects that enable novel extra functionality in GLSs. We first show that GRIN lenses (and, furthermore, GRIN cascades) can generate a multitude of vector vortex beams, and permit the creation of corresponding vector foci. We also demonstrate that, by proper choice of the interstitial optics of the cascade, we can use azimuthal or radial polarization light fields at the GRIN lens input to enhance the axial resolution. Lastly, we show how a GRIN cascade forms the basis of a novel single-shot Müller matrix polarimeter, which can comprehensively extract the polarization characteristics of an object. This capability can facilitate GLS based label-free endoscopic diagnosis of structural changes in tissue, with clear applications to cancer detection.

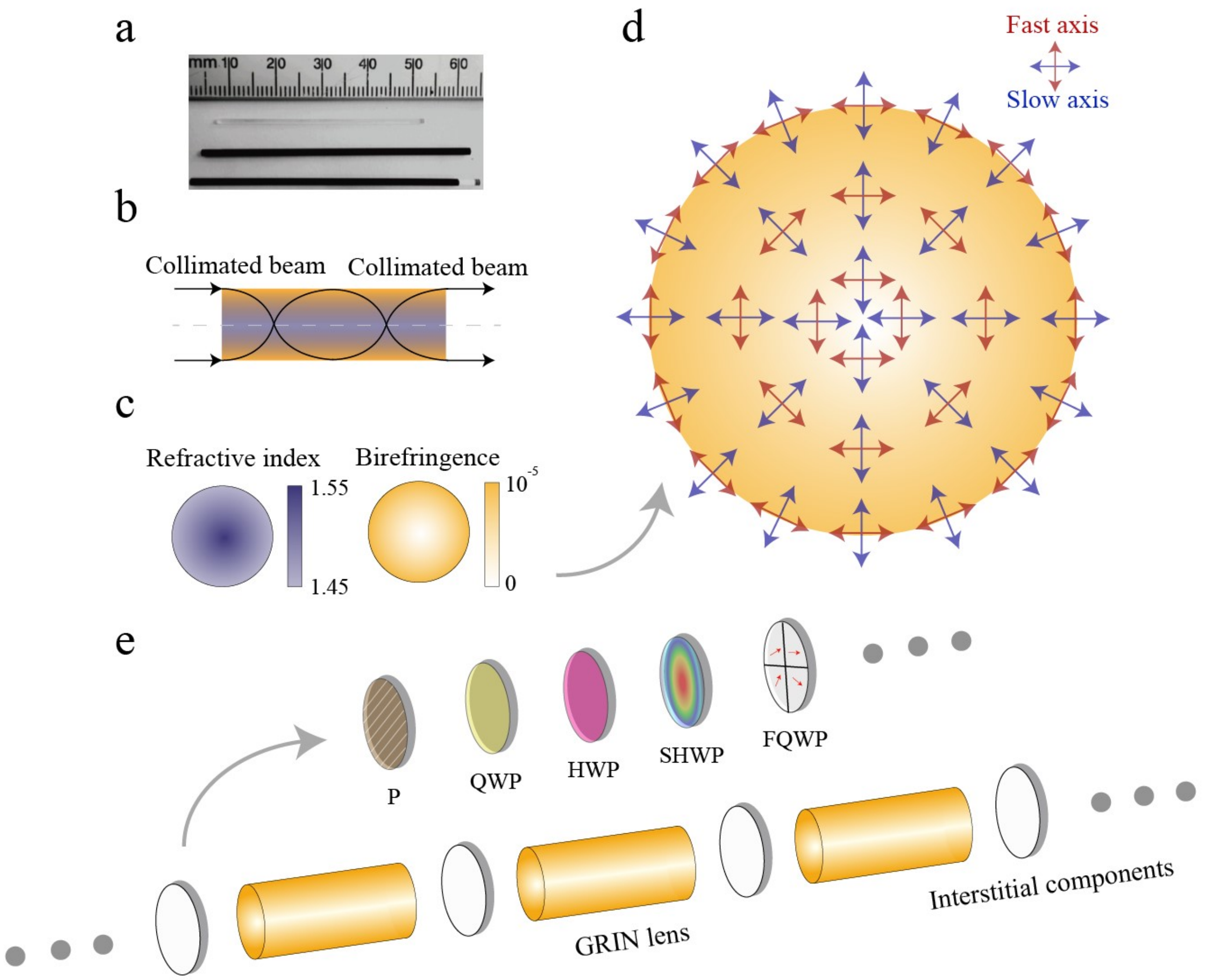

**Figure 1. GRIN lens properties and GRIN lens cascade.** (**a**) Commercial GRIN lenses. (**b**) Schematic of the GRIN lens ray trace. (**c**) The birefringence and refractive index profile of a GRIN lens over its cross section. (**d**) The fast axis/slow axis distribution of the local retardation across the GRIN lens. (**e**) The GRIN lens cascade. This cascade consists of the combination of one or more GRIN lenses along with various devices including interstitial components – such as P: Polarizer, QWP: Quarter waveplate; HWP: Half waveplate; SHWP: Spatially-variant half waveplate; FQWP: Four quadrant quarter waveplate array with four different fast axis orientations, etc.

## Results

### Vectorial beam generation by GRIN lens cascades

Great interest has developed in vector light fields that contain complex phase and polarization structures, whether in the focal or pupil domains. Of especial interest are those containing singularities, as they promise unprecedented capabilities for applications ranging from classical to quantum optics[11-17]. Examples include various complex vector beams that have non-uniform polarization distribution across the transverse plane[18-22] or beams with helicoidal wave fronts that carry orbital angular momentum (OAM)[23-31], or indeed their combinations as vector vortex beams (VVB)[32, 33]. Currently, such beams can be generated by using a range of modulation devices, such as spatial light modulators, q-plates[34], metasurfaces[35], segmented waveplates[36] or Fresnel-reflected cones[33]. We present here further options enabled by the birefringence of GRIN lenses especially for their usage in GLSs. The focussing properties of the GRIN lens arise due to the refractive index led dynamic phase that accounts for the sinusoidal ray trace using scalar theory. However, the birefringence means that vector effects must be considered. Furthermore, the lens is capable of manipulating the light through the introduction of a geometric Pancharatnam-Berry phase (PB-phase)[37]. To the best of our knowledge, this is the first time that the PB-phase and its corresponding effects have been considered for GLSs.

We first demonstrated VVB generation ability by using a single GRIN lens cascade. Examples of beam generation are shown in Fig. 2a through simulation and experiment, which show close correspondence (see methods in **Supplementary Note 2, 4 and 6**). It can be found from the polarization pattern that the beam is a specific singular vector beam, known as a full Poincaré beam[38, 39], that contains polarization singularities (see mathematical validation and polarization singular analysis in **Supplementary Note 4**). An interesting consequence of this is that the beam contains components of OAM, which have arisen due to the PB-phase (see theory in **Supplementary Note 6**). This was verified using a polarization state analyser (PSA) to select the left hand circular polarized state, and followed the commonly-used practice of using interference patterns[33, 35] (see method in **Supplementary Note 5**); the existence of two spirals validated that the single GRIN lens cascade could generate two units of OAM (see Fig. 2a iii, iv and v).

This concept can be extended further. Higher order GRIN lens cascades – those containing more than one GRIN lens, which may or may not be the same – are able to create even more complex VVBs, as shown in Fig. 2b to 2g. These results reveal that higher order cascades can generate complex polarization fields that have variations in radial or azimuthal directions, or encompass multiple sub-beams, each of which constitutes a full Poincaré beam; each of these Poincaré beams could have different topological charge or handedness, as well as beam profile (see analysis in **Supplementary Note 4**). The beams also contain different complex phase profiles; in **Supplementary Note 6** we illustrated further this phase manipulation ability.

It is important to realise that the spatial modulation effects described here are not confined to controlling the beam profile for the corresponding GLSs. They also have applications in focal shaping (as follows in section "Point spread function modulation" and the discussion), such as in fluorescence based GLS micro-endoscope, and application in GLS based polarization detection (as explained in section "Vectorial beam analysis" and the discussion), which could enable label-free sensing.

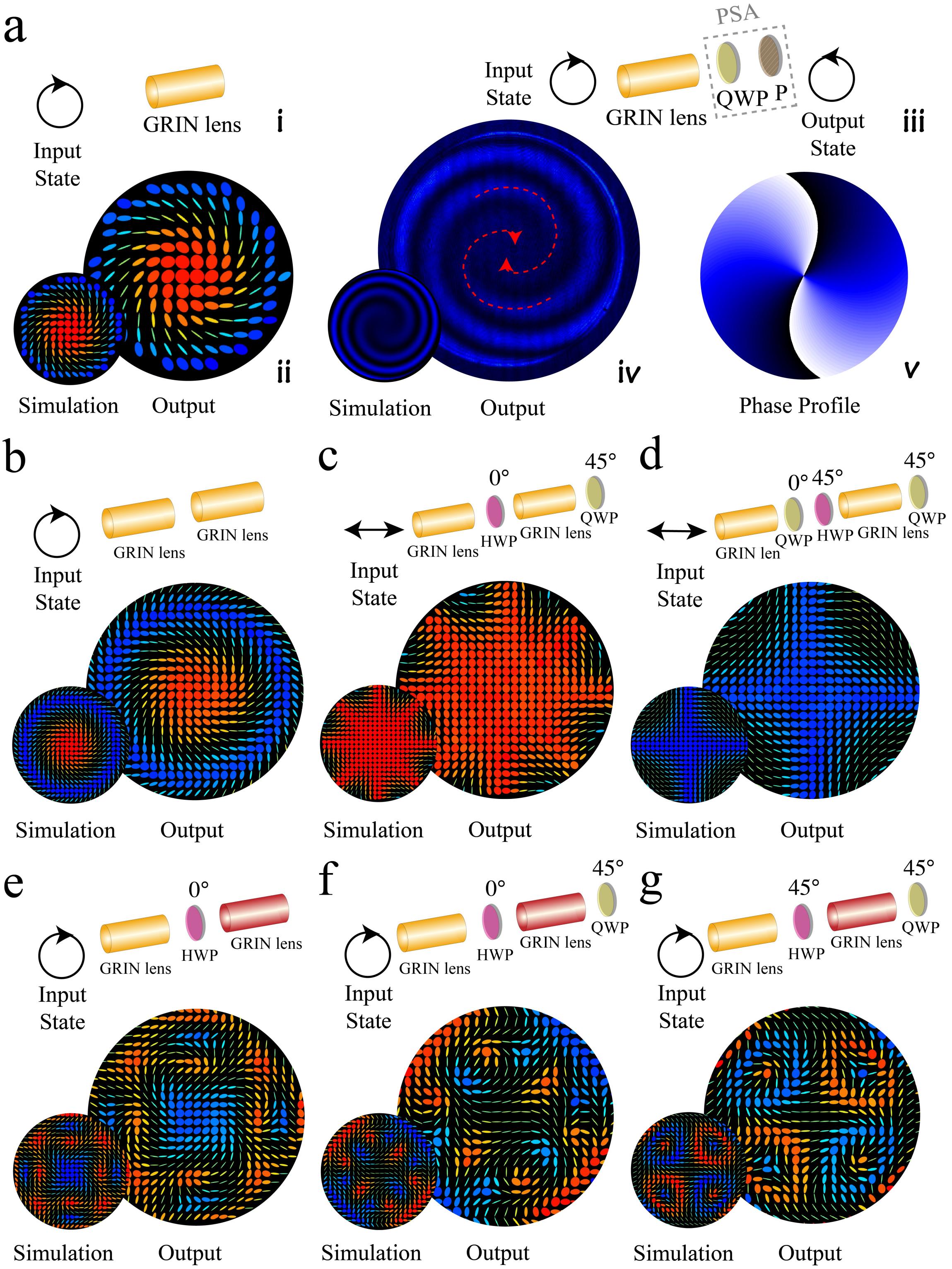

**Figure. 2 VVBs generation through GRIN lens cascades.** (**a**) (i) Schematic of a single GRIN lens cascade with (ii) right hand circular input, output polarization fields for both simulation and experimental results (the red side of the colour scale indicates right hand circular, whereas blue corresponds to left hand circular). Note that before the GRIN lens cascade there is a polarization state generator (PSG) to generate a chosen arbitrary polarization state; following the GRIN lens is a polarization state analyzer (PSA) that enables Stokes vector measurement (see method in Supplementary Note 3). Except in (iii) the PSG and PSA parts are omitted for clarity. (iii) An illustration for schematic in (a) with right hand circular input and left hand circular analysis. The two spirals in the interferogram (iv) and the phase profile (v) indicated that the light beam contained two units of OAM. (**b**), (**c**), (**d**), (**e**), (**f**), and (**g**) show higher-order GRIN lens cascades and their generated light fields under specific inputs. In (e), (f) and (g) the red-coloured GRIN lens indicates another GRIN type with a larger magnitude retardance profile almost equivalent to a sequence of three of the first (orange) type of lens (for details see Supplementary Note 4).

## Point spread function modulation by GRIN lens cascades

Due to the inherent link between the beam profile and the focus, polarization and/or phase modulation via the cascade also affects the point spread function (PSF) of a GLS. As beam modulations can also be harnessed in many imaging methods to enhance spatial resolution or other imaging properties[19, 20, 40], GRIN lens cascades have expected benefits in various GLSs. Indeed, we can show that through understanding these previously unwanted birefringence phenomena, it is possible to improve the performance of current GLSs, specifically as used for scanning micro-endoscopy.

Previous research has shown that GLS used in biomedical imaging can suffer from a lower than expected axial resolution; normally this is attributed to spherical aberration[41-43]. This has been a particular problem when using multiple pitch or long GRIN lenses, where the PSF has shown significant elongation[8]. We show here that this problematic phenomenon can be also introduced by the inherent birefringence of the GRIN lens (via the induced spatial polarization aberration). Furthermore, through understanding of these phenomena, we present a possible solution based upon specific polarization light fields that improves the axial resolution.

The fast and slow axis distributions of the GRIN lens (Fig. 1d) imply that the GRIN lens has two polarization eigenmodes, for which the beams would maintain the same polarization profiles throughout the device. These two modes – which are azimuthal and radial linearly polarized light fields – experience their own refractive index profiles $n_o(r)$ and $n_e(r)$, which are still approximately quadratic with $r$, but have different magnitudes. Hence, their focussing strength is different and the corresponding focussing pitch within the GRIN lens is different. As a consequence, the positions of the two corresponding foci are axially shifted relative to each other in the focal region (for the detailed mechanism see **Supplementary Note 2**). For this reason, other input states that are a combination of these eigenmodes – including uniform polarization states – create a superposition of both axially offset foci that lead to the elongated focus. However, this problem can be avoided if a single eigenmode is used.

To verify this proposition experimentally, we first measured the focus of GRIN lens when the input was a mixture of the two eigenmodes, using uniform circular polarization at the input (Fig 3a). The focal spot was elongated along the axis, with full width at half maximum (FWHM) of 17.4 μm (see **Supplementary Note 2** for the details of GRIN lens and methods). The inclusion of a spatially-variant half waveplate (SHWP) into the cascade, combined with appropriate input polarization, permitted generation of the two eigenmodes. The pupil fields at the output, as well as the axial PSF, were measured. We can see in Fig. 3b that the FWHM decreased to 14.1 μm and, in Fig. 3c, that a focal shift of 3.0 μm was present between the foci of the two eigenmodes, confirming our theory. Note that the foci for the individual eigenmodes were ring shaped in the lateral plane, with zero intensity along the axis due to the phase singularity introduced by the SHWP in the cascade. In a practical imaging system, this effect could be compensated using an appropriate helicoidal phase plate[44], or an alternative method for preparation of the input polarization state. Our observations show that enhanced axial resolution can be obtained in a GLS by modulation of the focus through the appropriate configuration of the input polarization and phase state.

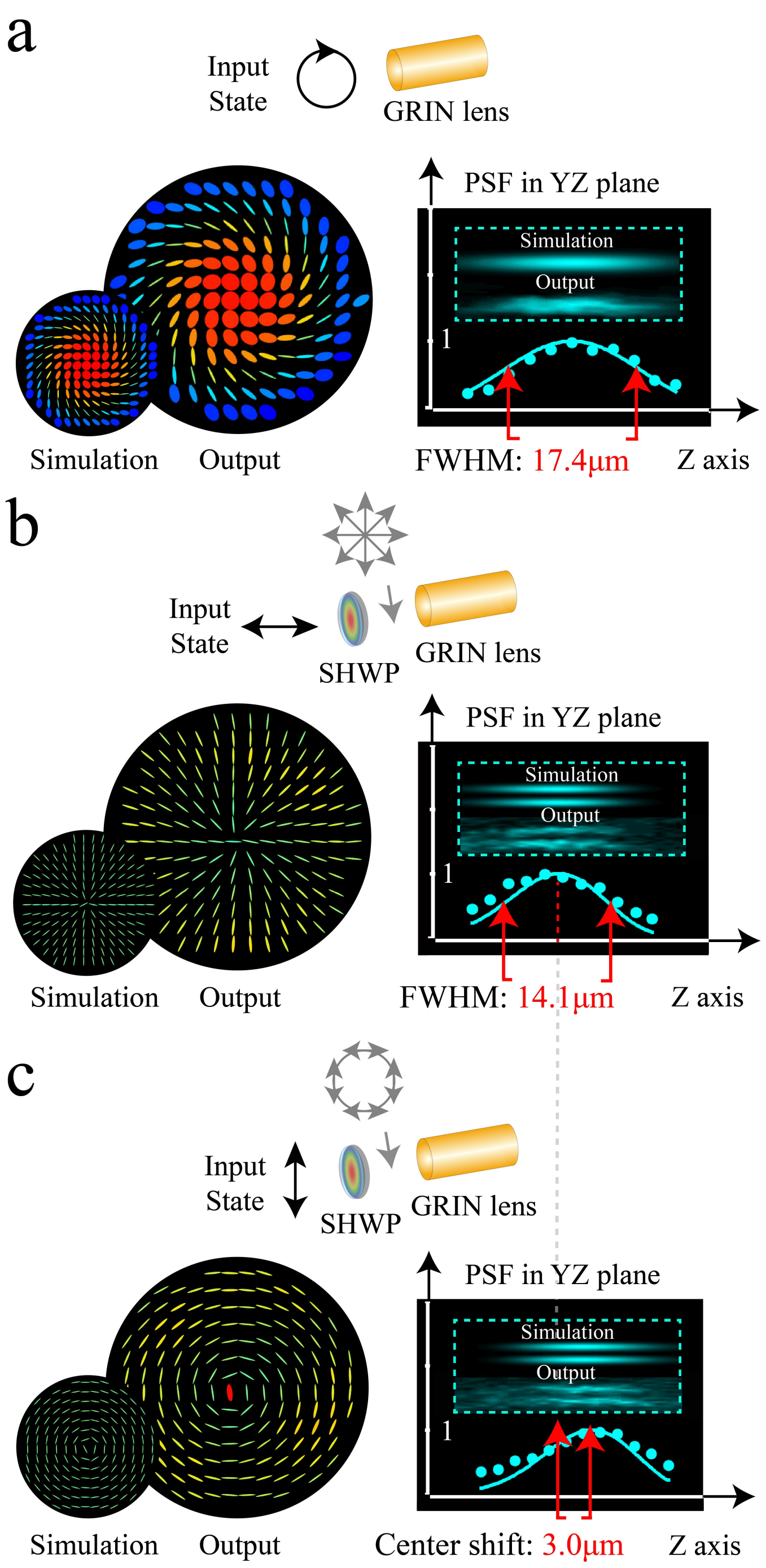

**Figure. 3 PSF modulation through GRIN lens cascade.** (**a**) Schematic of a single GRIN lens cascade illuminated by right hand circular polarized light, beam profile after the GRIN lens, and the corresponding axial PSF including the FWHM. (**b**) and (**c**) show the GRIN lens cascade including SHWP with horizontal and vertical linear polarized light input, beam profile after the GRIN lens, and the corresponding axial PSF to demonstrate the focus shift when using the radial/azimuthal polarization input light fields.

**Vectorial beam analysis through GRIN lens cascades**

As shown in the previous section, there is a complex interdependence of the focal field and the polarization state of the input beam, as processed through the GRIN lens cascade. This dependence provides further opportunities to harness the birefringence properties of the GRIN lens as parallel analysis channels for polarization sensing. The primary ability of the GRIN lens as a focussing device is due to its continuously varying, rotationally symmetric refractive index distribution[9, 10]. As already discussed, this is accompanied by the gradually varying birefringence distribution. As a high degree of symmetry is required for focusing, this means that the birefringence distribution is also highly symmetric. This provides a continuous, non-pixellated, range of analysis channels that could be used for polarimetry.

Müller matrix (MM) polarimetry is a commonly used method to measure the full polarization properties of materials, particularly in biomedical applications, where its ability to characterise tissue structures in a label-free manner is highly advantageous[45-55]. Numerous MM polarimeters have been proposed, many of which are based on sequential generation of states of polarization (SOP) and analysis by rotation of polarization components or modulation of variable retarders[48]. However, in general, sequential measurement is not suitable for fast detection, such as for moving objects where it leads to unexpected measurement errors[51]. The speed can also limit the application of MM polarimetry to *in vivo* detection of a dynamic sample, such as in clinical applications.

We propose here a new compact single-shot MM polarimeter based on a GLS – specifically, a dual GRIN lens cascade – which can directly act as a basis of rigid endoscopic system. It includes a PSG consisting of a polarizer and a four-quadrant quarter waveplate (Fig. 4b (i)) to generate simultaneously four different illumination SOPs. Figure 4b (ii) shows the optimized SOPs on the Poincaré sphere as generated by the PSG (see design details in **Supplementary Note 7**). The GRIN lens cascade-based PSA is shown in Fig. 4a; here, a half-wave plate was sandwiched between two identical GRIN lenses, which were followed by a polarizer. We applied the Lu-Chipman MM polar decomposition (MMPD)[56] method (see method in **Supplementary Note 1**) to assess the equivalent retardance and fast axis direction of the PSA – these are shown for both simulation and experiment in Fig. 4c (ii) and (iii). Each of the four quadrants of the PSA provides access to a continuum of polarization analysis channels. Taking advantage of these channels, the GRIN lens cascade can be employed as a complete analyser to calculate four Stokes vectors in a single-shot – one for each of the four channels, which thus enables the single-shot MM measurement (see details of characterisation in **Supplementary Note 8 and 9**).

To illustrate the potential of the GLS polarimetry for clinical diagnosis, the system was tested on tissue samples. We measured the MMs of the unstained 12-μm-thick human breast ductal carcinoma tissue samples at two stages (healthy vs. invasive carcinoma) and we derived the retardance parameter from the MMs to permit quantitative comparison (see methods in **Supplementary Note 1**). Clinically, different stages of breast ductal carcinoma tissues have different proportions and distribution of fibrous structures in and around the milk ducts[52]. An important potential application of MM polarimetry is to distinguish between such stages of cancerous tissue in a label-free manner[52]. We first demonstrated the method on one healthy tissue sample (named Sample H1, where H represents healthy) and one carcinoma tissue sample (named Sample C1, where C represents cancerous). The number of the random sampled points measured by the GRIN lens cascade for each was fixed at 10 throughout this work (see Fig. 4d, sample information and corresponding methods in **Supplementary Note 10**). Figure 4e and 4f reveal the statistical distribution of data measured by the cascade and MM microscope, as ground truth (see data details and the corresponding P-value analysis in **Supplementary Note 10** and **Supplementary Table 1**). We found from the data that the measured values of retardance for Sample C1 were significantly larger than those of the Sample H1 just as the same significant difference existed among the data measured by MM microscope. Furthermore, the measured values from all 10 breast ductal carcinoma samples ($1324 \pm 433$ rad/$10^4$) were significantly larger than those of the 10 healthy ones ($408 \pm 189$ rad/$10^4$) – these retardance measurements were calculated from the original measured data (in Fig. 4f) through standard statistical methods[52] to obtain their mean values and standard deviations (see details in **Supplementary Note 10**). The difference is due to the larger proportion of fibrous structures present in the cancerous tissues, which is a consequence of the inflammatory reactions induced by cancer cells[52]. The results reveal that the GRIN lens cascade based polarimeter has the ability to differentiate between healthy and cancerous breast tissues.

These experiments show that the GRIN cascades can take advantage of the previously unwanted birefringence not only to create vector foci and improve axial resolution in the focus, but also to enable polarimetric measurement. This has potential to act as a single-shot, compact, polarization detection probe to assist minimally invasive clinical diagnosis.

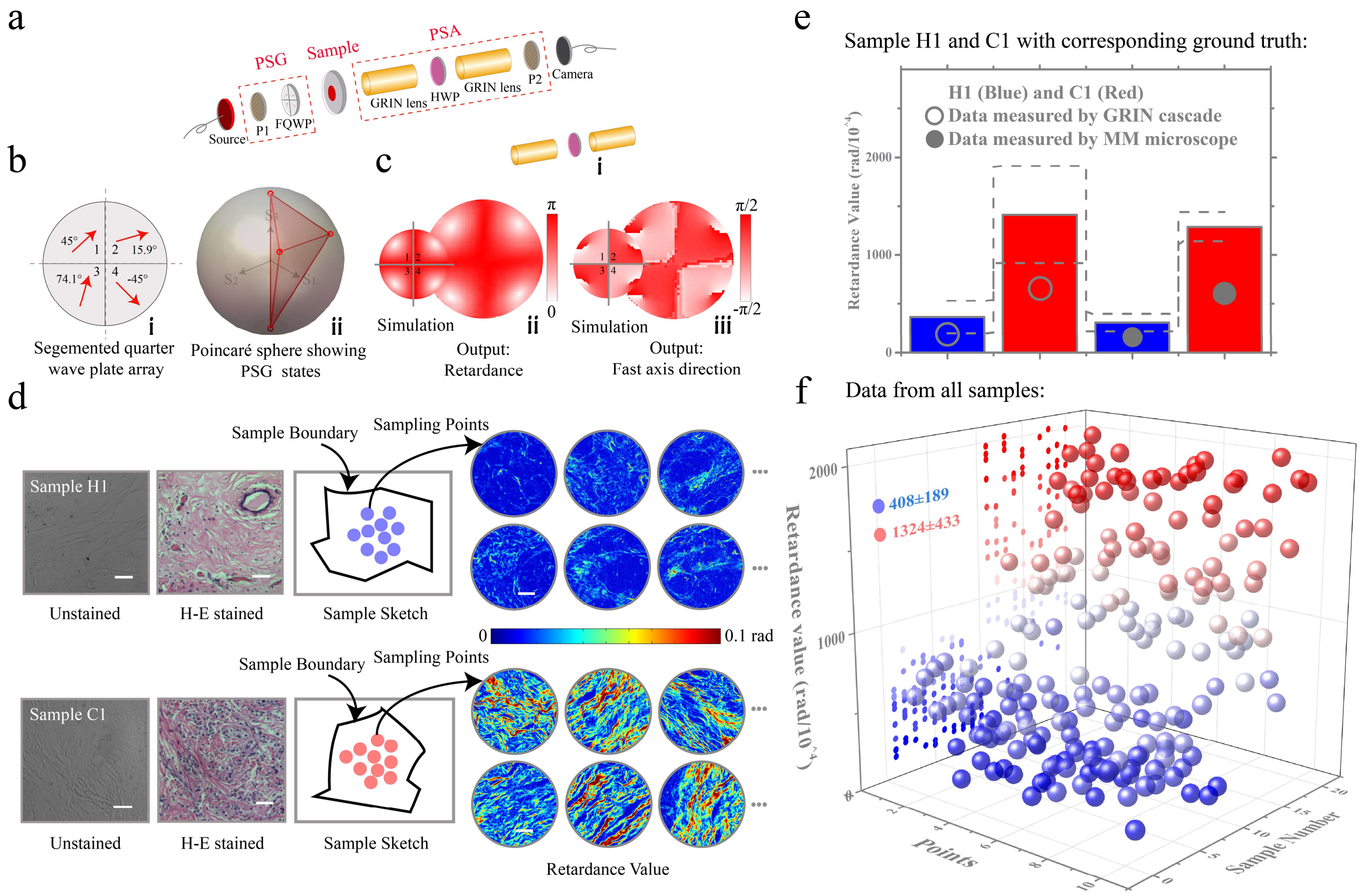

**Figure. 4 Characterisation of the GRIN lens based MM polarimeter and measurement of tissue samples.** (**a**) Simplified setup of the polarimeter. PSG: P1, polarizer; FQWP, four-quadrant quarter waveplate, with different fast axis orientations in each quadrant. PSA: P2, polarizer; HWP, half waveplate. (**b**) (i) and (ii) details of FQWP and the Poincaré sphere showing the SOPs generated by the PSG. (**c**) (i) The GRIN lens cascade in the PSA besides the P2. (ii) and (iii) Simulation and experimental results of retardance and the fast axis orientation of (i), derived by the MMPD method. (**d**) Demonstration of the samples H1 and C1 (unstained samples and their H-E stained counterparts), sketches with corresponding random sampling points (see method in Supplementary Note 10) and examples of retardance distributions measured by the MM microscope (as ground truth). Scalar bar: 50µm. (**e**) Statistic histogram (mean value and the standard deviation of the retardance) of the 10 points from each sample within the same region, measured by the GRIN lens cascade and a conventional MM microscope. Standard deviations are shown via the dashed lines. Numbers refer to Supplementary Table 1. Source data are provided as a Source Data file. (**f**) Overall data distribution (retardance value). The overall number of samples was 20 (half cancerous *vs*. half healthy, giving 10 testing points per sample). Note that blue (ball or histogram) represents healthy samples while red represents cancerous samples.

**Discussion**

By understanding and harnessing the inherent – and previously undesirable – birefringence of GRIN lenses, we have opened up new possibilities for their use. Combining one or more GRIN lenses into cascades with other optical elements, provides a wide range of complex beam modulation capabilities to benefit current GLSs, including vector vortex beam generation, focal spot shaping, and in polarization analysis. All of those demonstrations are achieved using standard, off-the-shelf optical components and compatible with the GRIN lens' basic functions as a focusing and imaging device.

The fundamental realizations here about the joint effects of the intrinsic birefringence and the focusing properties of the GRIN lens are also important in understanding some of the current challenges in use of various GLSs[1-8]. Our observations on the dimensions of the pupil/focal distributions, the use of polarization eigenmodes, and the improvement of axial resolution, show that better understanding of their vector modulation effects can benefit current GLSs, such as mitigation against aberration problems in GRIN lens based two-photon fluorescence micro-endoscopy. Furthermore, the well-controlled creation of ring-shaped foci through vectorial beam manipulation could now enable further lateral resolution enhancements for GLS through stimulated emission depletion[19, 20] (STED), MINFLUX[57] or related microscopy methods. These methods are particularly sensitive to polarization and phase aberrations, so improved understanding of the birefringence effects and their mitigation will be essential if they are to be combined with GLSs. This may find application in *in vivo* deep tissue (such as brain) imaging, which cannot be easily achieved by alternative objective lenses. Combination of the GRIN cascades with active optical elements[58] in cascade, such as a spatial light modulators, could enable further applications in beam or focal control. Further imaging methods for GLSs are then expected across various existing techniques such as structured illumination microscopy[59] (SIM).

Our observations that the GRIN cascades perform optically efficient, continuous, non-pixellated, vectorial beam modulation have important consequences for many applications. The geometric phase[16] effects generated within the GRIN cascades are more complex than those from planar wave plate arrays, not only in the spatial domain but also in the time domain (see details in **Supplementary Note 2**), which may open up interesting prospects for quantum state manipulation. The observation of OAM generation could also have potential for further GRIN lens based micro-manipulation applications, which take advantage of the optical torques introduced by the phase vortex. As certain cascades show highly asymmetric MMs (see details in **Supplementary Note 1**), there is scope for further investigation of these unusual properties. The same cascades also exhibit gradients in circular anisotropy, which presents the intriguing possibility that the spin-Hall effect of light could be harnessed in a GRIN lens based imaging system (see details in **Supplementary Note 1**). Perhaps the most immediate application of this new concept is in polarimetry. The GRIN cascade provides the basis a miniaturizable, cost-effective, compact and multifunctional probe with the capability for (endoscopic[55]) MM polarimetric diagnosis in the clinical setting and it also has the potential for widefield or scanned imaging. The demonstrations in this article were performed at a single wavelength, but could be expanded to multiple wavelength channel polarimetry, while still maintaining optimal MM measurement for each wavelength.

Overall, the optical properties of GRIN lenses and their combination in cascade structures provide a wealth of opportunity for further development. Future benefits are expected in a wide range of applications, spanning from quantum optics to clinical diagnosis.

## Acknowledgments

We are grateful to Dr. Neal Radwell from University of Glasgow, Dr. C. T. Samlan from University of Hyderabad, Miss. Yang Dong, Mr. Donghong Lv, Prof. Yonghong He and Prof. Nan Zeng from Tsinghua University, Prof. Hailu Luo from Hunan University, Prof. Desen Liu, Prof. Xiaoping Jiang and Prof. Sumei Zhou from Southwest University, Dr. Ji Qi from University College London, Dr. Patrick Salter from University of Oxford, Mr. Zhao Liu from Femto Technology Co. Ltd., and Mr. Herbert Stürmer from GRINTECH GmbH for the discussions about various topics. The project is supported by the European Research Council (AdOMiS, no. 695140).

## Author contributions

C.H. and M.B. conceived the main ideas for the applications of GRIN lens cascades, developed the concepts and planned the modelling and experiments. J.C., C.H., J.W., Q.H. and J.A. performed GRIN lens equations, contributed to polarization measurement methods and interferometry. B.D., Q.H., C.H., J.C., J.L. and P.X. performed the simulations, VVB mathematical approaches and the analysis of OAM components. J.W., C.H. and H.H. performed the PSF measurement and tested the MM polarimeter, S.L. prepared the samples. J.C., J.L., H.H., C.H., D.E. and H.M. contributed to MMPD method, MM calculation algorithm and MM analysis method. C.H. carried out the remaining experiments, prepared the figures, and analyzed the results. C.H. and M.B. wrote the paper. M.B. oversaw and led the project. All authors participated in discussion and made comments for the paper.

## Additional Information

Correspondence and request for materials should be addressed to C.H. or M.B. The use of the human tissue slices was approved by the Ethics Committee of the Shenzhen Sixth People's (Nanshan) Hospital.

## Supplementary Information

Supplementary Information accompanies this paper at http://www.nature.com/naturecommunications

**Data availability**

Data generated and analysed during this study are included in this article and its Supplementary Information files, and are also available from the corresponding authors on reasonable request. The source data underlying Fig. 4e and Supplementary Figure 11c are provided as a Source Data file.

**Competing interests**

The authors declare no competing interests.

# Complex vectorial optics through gradient index lens cascades

## – Supplementary Information –

He et al.

## Supplementary Notes

Chao He[1,*], Jintao Chang[2,3], Qi Hu[1], Jingyu Wang[1], Jacopo Antonello[1], Honghui He[3], Shaoxiong Liu[4], Jianyu Lin[5], Ben Dai[6], Daniel S. Elson[5], Peng Xi[7], Hui Ma[2,3] and Martin J. Booth[1,*]

[1]*Department of Engineering Science, University of Oxford, Parks Road, Oxford, OX1 3PJ, UK*

[2]*Department of Physics, Tsinghua University, Beijing 100084, China*

[3]*Shenzhen Key Laboratory for Minimal Invasive Medical Technologies, Institute of Optical Imaging and Sensing, Graduate School at Shenzhen, Tsinghua University, Shenzhen 518055, China*

[4]*Shenzhen Sixth People's Hospital (Nanshan Hospital) Huazhong University of Science and Technology Union Shenzhen Hospital, Shenzhen 518052, China*

[5]*Hamlyn Centre for Robotic Surgery, Institute of Global Health Innovation, Imperial College London, London SW7 2AZ, UK*

[6]*School of Data Science, City University of Hong Kong, Kowloon, Hong Kong, China*

[7]*Department of Biomedical Engineering, College of Engineering, Peking University, Beijing 100871, China*

[*]*Corresponding authors:* chao.he@eng.ox.ac.uk; martin.booth@eng.ox.ac.uk

## Supplementary Note 1: Characterization of polarization properties of the GRIN lens

The polarization properties of an object can be described using either a Jones matrix (JM) or Müller matrix (MM)[1, 2]. While Jones matrices are used to deal with interference cases by fully polarized light, the MM contains 16 elements $m_{kl}$ $(\mathrm{k},\mathrm{l} = 1,2,3,4)$, which can comprehensively represent the polarization characteristics – including depolarization – of the target[1, 2]. In this section, we use the MM to characterize the polarization properties of GRIN lens cascades. These measurements use a calibrated MM imaging polarimeter based upon the dual rotating waveplates method[3-5]. The measurement principle and schematic of the setup are shown in the following contexts.

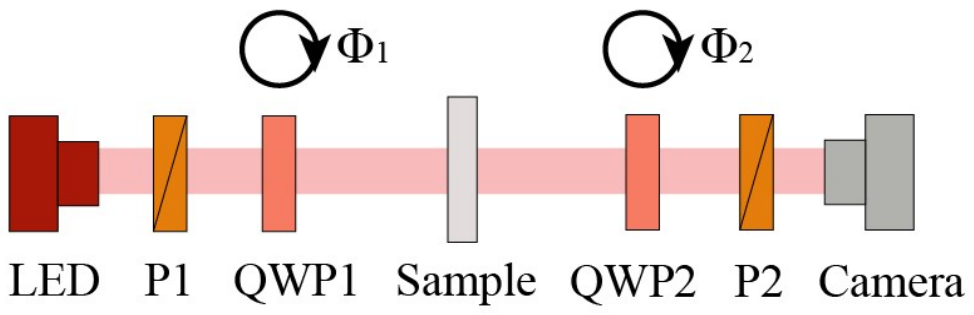

**Supplementary Figure 1. Setup of the MM imaging polarimetry.** LED light source; P1, P2: fixed polarizer; QWP1, QWP2: rotated quarter waveplate. Camera as detector.

Supplementary Figure 1 shows the MM imaging polarimeter. The polarizers (P1, P2) are oriented in the same direction. Two quarter waveplates (QWP1, QWP2) rotate with fixed rotational speeds, such that $\phi_1 = 5\phi_2$. The main measurement principle is shown in Eq. (1) below, where superscript $q$ represents the $q^{\mathrm{th}}$ measurement. $M_{\mathrm{Sampl}}$ is the MM of the sample, $M_{\mathrm{P1}}$, $M_{\mathrm{P2}}$, $M_{\mathrm{QWP1}}$, $M_{\mathrm{QWP2}}$ are MMs of P1, P2, QWP1, QWP2, respectively. $M_{\mathrm{System}}$ is the equivalent overall MM of the system. $S_{\mathrm{in}}$ and $S_{\mathrm{out}}$ are incident and output Stokes vectors.

$$S_{\mathrm{out}}^{q} = M_{\mathrm{System}} S_{\mathrm{in}} = M_{\mathrm{P2}} M_{\mathrm{QWP2}} M_{\mathrm{QWP2}}^{q} M_{\mathrm{Sampl}}\ M_{\mathrm{QWP1}}^{q} M_{\mathrm{P1}} S_{\mathrm{in}}. \quad (1)$$

Since the intensity is equivalent to the first element $S_0$ of the Stokes vector, we make $I^q = (S_{\mathrm{out}}^{q})_0$, which represents the corresponding intensity of $q^{\mathrm{th}}$ measurement. From Eq. (1) we obtain the Fourier series,

$$I^q = (S_{\mathrm{out}}^{q})_0 = a_0 + \sum_{n=1}^{12} (a_n \cos 2n\phi_1^q + b_n \sin 2n\phi_1^q), \quad (2)$$

where $a_n$ and $b_n$ are the Fourier coefficients, and $\phi_1^q$ is the angle of QWP1 at $q^{\mathrm{th}}$ measurement. The MM of the sample could then be calculated from the Fourier coefficients as shown below. For more details see Ref [2-4].

$$M_{\text{Sample}} = \begin{pmatrix} m_{11} & m_{12} & m_{13} & m_{14} \\ m_{21} & m_{22} & m_{23} & m_{24} \\ m_{31} & m_{32} & m_{33} & m_{34} \\ m_{41} & m_{42} & m_{43} & m_{44} \end{pmatrix} =$$

$$\begin{pmatrix} a_0 - a_2 + a_8 - a_{10} + a_{12} & 2a_2 - 2a_8 - 2a_{12} & 2b_2 - 2b_8 - 2b_{12} & b_1 - b_9 - b_{11} \\ -2a_8 + 2a_{10} - 2a_{12} & 4a_8 + 4a_{12} & 4b_{12} - 4b_8 & -2b_9 + 2b_{11} \\ -2b_8 + 2b_{10} - 2b_{12} & 4a_8 + 4b_{12} & 4a_8 - 4a_{12} & 2a_9 - 2a_{11} \\ b_3 - b_5 + b_7 & -2b_3 - 2b_7 & -2a_3 + 2a_7 & -a_4 + a_6 \end{pmatrix}. \quad (3)$$

Once the MM of the sample is obtained, we apply the Lu-Chipman MM polar decomposition method (MMPD)[6] to extract polarization parameters from the MM. This method is widely used to decompose the complicated interactions between sample and polarized light into a series of phenomena[6-10]: the sample's diattenuation ($D$), depolarization ($\Delta$), retardance ($R$) and its fast axis direction ($\theta$) represented by the corresponding three Matrix factors $M_\Delta$, $M_R$, $M_D$, respectively. The main principle is represented by Eq. (4).

$$M_{\text{Sample}} = M_\Delta M_R M_D. \quad (4)$$

The diattenuation value $D$ can be readily obtained from the second to fourth elements in the first row of a MM, that is the elements $m_{12}$, $m_{13}$ and $m_{14}$ respectively, as shown in Eq. (5). The retardance R is reconstructed from the trace of $M_R$, Eq. (6), while the orientation of optic axis ranging from $-\frac{\pi}{2}$ to $\frac{\pi}{2}$ radians is calculated according to Eq. (7). The depolarization properties are included in the in the bottom right 3×3 matrix $m_\Delta$ of the matrix $M_\Delta$, which is shown in Eq. (8). $\lambda_1$, $\lambda_2$ and $\lambda_3$ in Eq. (8) are the eigenvalues of $m_\Delta$, and $P$ is a matrix composed of the eigenvectors of $m_\Delta$. $\Delta_L$ and $\Delta_c$ in Eq. (8) are linear/circular depolarization values.

$$D = \sqrt{m_{12}^2 + m_{13}^2 + m_{14}^2}, \quad (5)$$

$$R = \cos^{-1}\left[\frac{\text{tr}(M_R)}{2} - 1\right], \quad (6)$$

$$\theta = \frac{1}{2}\tan^{-1}\left[\frac{M_{R2} \; - M_{R32}}{M_{R31} - M_{R13}}\right], \tag{7}$$

$$m_\Delta = P\begin{pmatrix} \lambda_1 & 0 & 0 \\ 0 & \lambda_2 & 0 \\ 0 & 0 & \lambda_3 \end{pmatrix}P^{-1}, \;\; \Delta_L = 1 - \frac{\lambda_1+\lambda_2}{2}, \;\; \Delta_R = 1 - \lambda_3. \tag{8}$$

Supplementary Figures 2a to 2c demonstrate the MMs and the corresponding MMPD results of the single GRIN lens cascade. We found that 1) the measured experimental MM as well as the MMPD parameters matched well with the simulation counterparts; 2) the MMPD results reveal that the birefringence structure of the GRIN lens behaves equivalently to a spatially-variant waveplate array with a) linear retardance that gradually increases along the radial direction and b) fast axis directions that vary azimuthally (gradually changing from -π/2 to 0 then to π/2 radians as the azimuthal angle changes from 0 to π radians, then a similar variation when the azimuthal angle changes from π to 2π radians). The fast/slow axis direction distributions are shown in Fig. 1b in the main article as well.

In Supplementary Figures 2d to 2g, we further show four different SHWP based GRIN lens cascades including in Supplementary Figure 2d – the one used in the main article (as Fig. 3b and 3c). Besides the close correspondence between simulation and experimental results, we also found that the MMs of these GRIN lens cascades have significant off-diagonal asymmetry which is of interest in MM research, as well as related sample information analysis[11]. Specifically, we extracted the circular anisotropy coefficient[12] of these four MMs, shown in Supplementary Figure 2h, in which we use grey arrows to indicate the gradient of the circular anisotropy[13]. This implies the system can be used to demonstrate spin-Hall effect (SHE) of light[13]. Both of these phenomena deserve further exploration.

We propose here that the GRIN lens cascade can also find use as a special device for other research areas, such as the derivation of quantitative MM parameters from exotic materials (e.g. chiral characteristics[14]), biomedical research (multi-layer birefringence and/or diattenuation characterization[11]), or studies on complex spin-orbital interaction (SOI) processes of light[13]. These all take advantage of the unique property of the GRIN lens – its equivalence to a spatially variant waveplate array, which can encompass all combinations of retardance and fast axis direction.

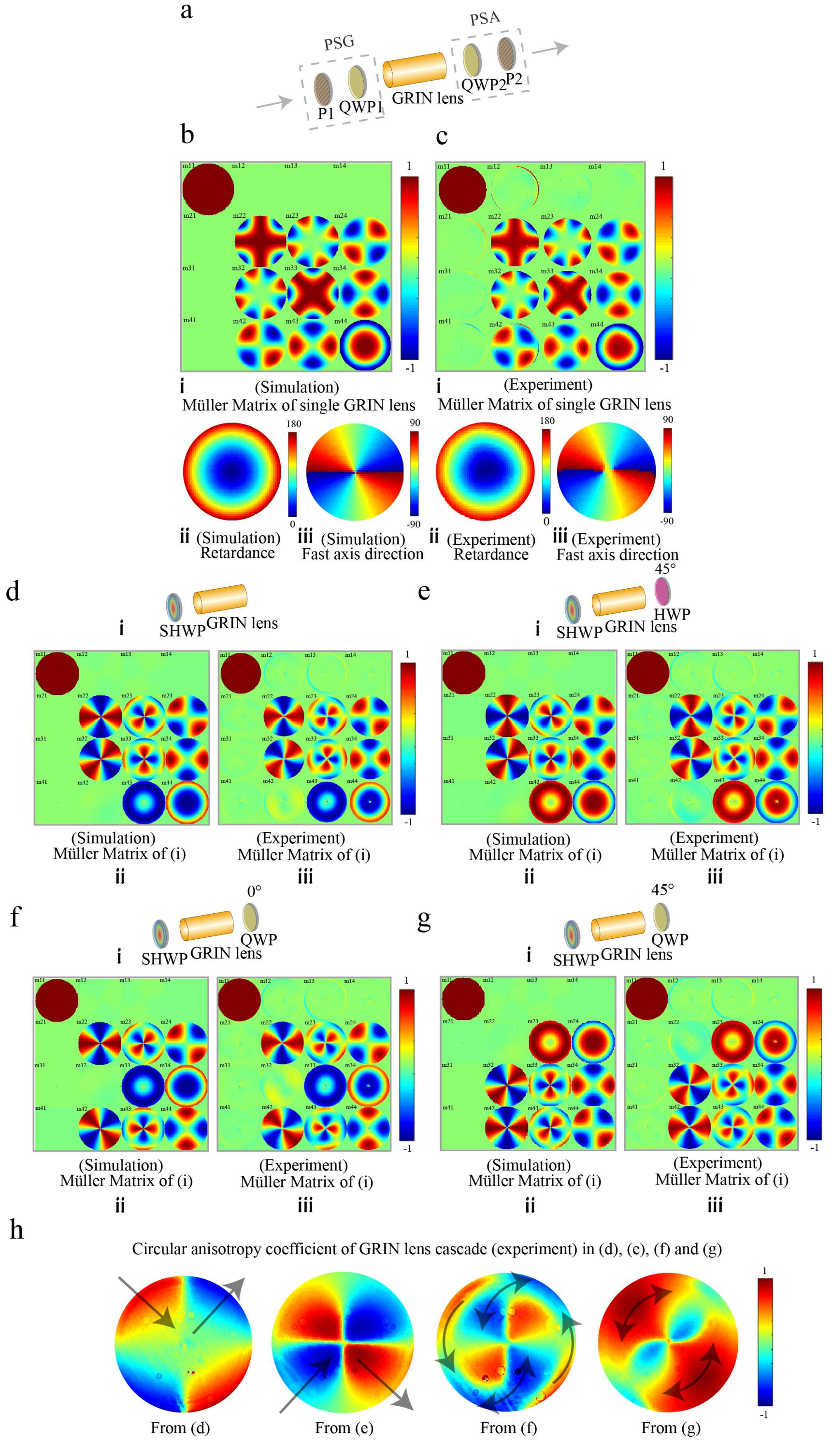

**Supplementary Figure 2. Polarization properties of several GRIN lens cascades.** (**a**), (**b**) and (**c**) are the schematic of the single GRIN lens cascade and its MM, polarization parameters decomposed by MMPD method both in simulation and experiment. (**d**), (**e**), (**f**), (**g**) are four different SHWP based GRIN lens cascades (the diagrams omit the polarization state generator (PSG) and a polarization state analyser (PSA) parts) with their simulated and experimental MMs. (**h**) Circular anisotropy coefficient of GRIN lens cascade in (**d**), (**e**), (**f**) and (**g**). The grey arrows indicate the anisotropy gradient.

## Supplementary Note 2: Modelling of light modulation by the graded birefringence of the GRIN lens

In this section we show mathematically how the retardance profile of the GRIN lens can be calculated and show how the propagation of vector fields through these lenses can be modelled (the equations in this section form the basis for all modelling of the GRIN cascades).

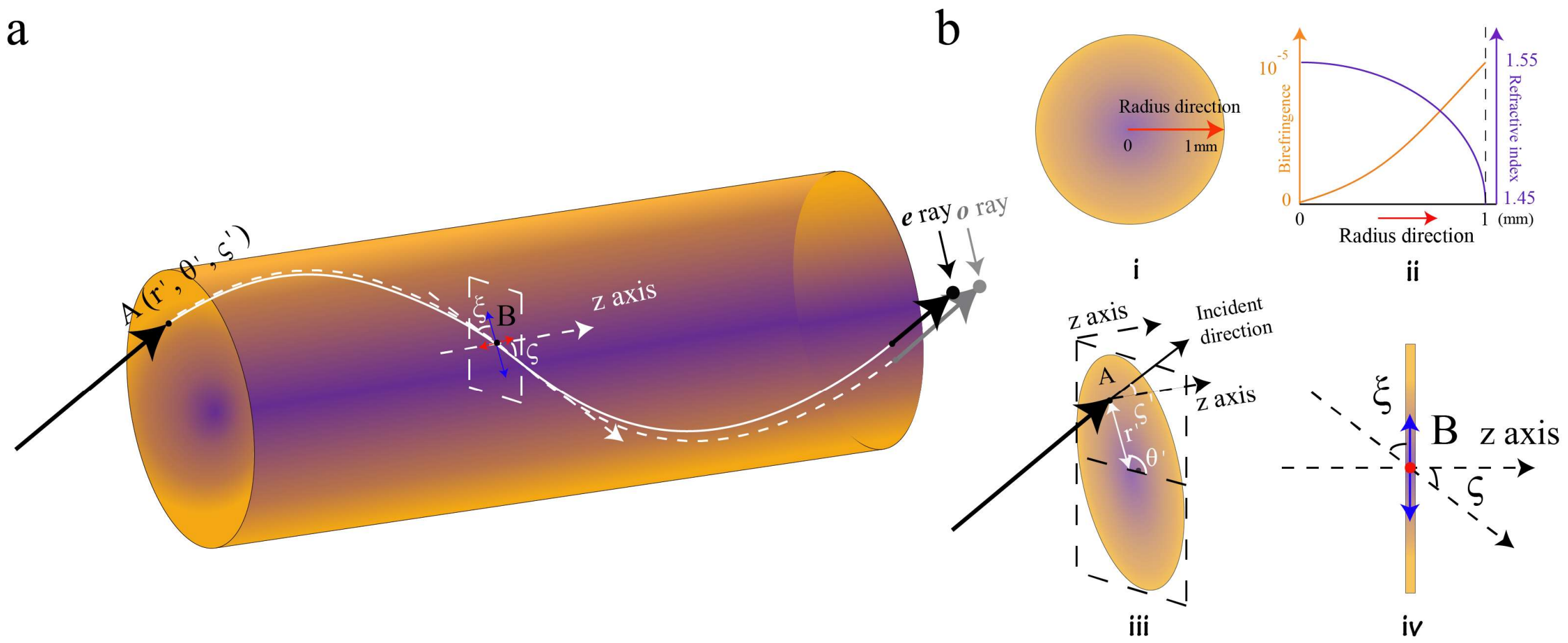

**Supplementary Figure 3. Model for analysing the retardance profile of a GRIN lens.** (**a**) A specific incident point A is represented by $(r', \theta', \varsigma')$, and the arbitrary point inside GRIN lens along the propagation ray trace is denoted by B $(r, \theta, \varsigma)$. The blue double-sided arrow is slow axis direction (extraordinary axis), and the orthometric red counterpart is the fast axis direction (ordinary axis) of that arbitrary point. The black and grey focus points at distal side of the GRIN lens represent focus splitting between the *e* ray (white solid line) and the *o* ray (white dotted line), which is introduced by both the birefringence property and the sinusoid ray trace. (**b**) (i) the section of a GRIN lens shows the refractive index and the birefringence profiles, (ii) shows two curves of these two parameters along the radius direction from centre point to the boundary of GRIN lens, (iii) shows an incident light on point A$(r', \theta', \varsigma')$, on the front surface of GRIN lens, referenced by z axis. (iv) shows the lateral profile of one arbitrary section of GRIN lens, which includes a ray trace passing through point B. $\xi$ is the interior angle between the ray and the extraordinary axis (blue double-sided arrow in (**a**)) at point B, which is the complementary angle to $\varsigma$, which is the interior angle between the ray and the direction of the z axis.

Supplementary Figure 3a illustrates the refractive index (violet coloured gradient), birefringence (yellow coloured gradient) as well as one arbitrary ray trace (white solid line) when light propagates through the GRIN lens. Supplementary Figure 3b (i) shows the refractive index profile and birefringence profile of a GRIN lens section, and (ii) shows sketches of these two parameters along the radius direction from the centre to the boundary (refer to the lens from Femto Technology Co. Ltd., G-B161157-S1484). All the calculations below are under the meridional plane approximation[16].

For a general ray within the GRIN lens, the distance from the GRIN lens axis is denoted by $r$, the azimuth of the incident position is denoted by $\theta$, and the interior angle between the ray and the direction of the $z$ axis is denoted by $\varsigma$. Then a specific incident ray at point A can be expressed in terms of $(r', \theta', \varsigma')$, also shown in Supplementary Figure 3b (iii). With traditional GRIN lens ray tracing[16], the ray path $C$ (which represents the radius at distance $z$) of the specific incident beam can be represented by:

$$C(z) = r' \cos(\sqrt{A}z) + \frac{\tan \varsigma'}{\sqrt{A}} \sin(\sqrt{A}z) = \sqrt{r'^2 + \frac{\tan^2 \varsigma'}{A}} \cos\left(\sqrt{A}z - \tan^{-1}\left[\frac{\tan \varsigma'}{\sqrt{A}r'}\right]\right), \quad (9)$$

where $z$ is the propagation length along the axis and $A$ is a constant determined by the manufacturing process that is directly related to the refractive index distribution, $\frac{2\pi}{\sqrt{A}}$ is the period of the sinusoidal ray trace and the amplitude is $\sqrt{r'^2 + \frac{\tan^2 \varsigma'}{A}}$. The refractive indices seen by the $o$ rays and $e$ rays are denoted by $n_{\mathrm{o}}$ and $n_{\mathrm{e}}$ respectively, which are functions of the radius $r$. We then could express the refractive index of a GRIN lens in series form to approximate the real distributions:

$$\begin{aligned} n_{\mathrm{o}}(r) &= n_{\mathrm{o}}(0) + \alpha_1 r + \alpha_2 r^2 + \alpha_3 r^3 + \cdots + \alpha_k r^k, \quad (k = 1,2,3,\cdots) \\ n_{\mathrm{e}}(r) &= n_{\mathrm{e}}(0) + \beta_1 r + \beta_2 r^2 + \beta_3 r^3 + \cdots + \beta_k r^k, \quad (k = 1,2,3,\cdots) \end{aligned} \quad (10)$$

Where $n_{\mathrm{o}}(0)$ and $n_{\mathrm{e}}(0)$ are the refractive indices of the $o$ rays and $e$ rays at the centre (note that $n_{\mathrm{o}}(0) = n_{\mathrm{e}}(0)$), and $\alpha_1, \alpha_2, \alpha_3, \cdots \alpha_k$ and $\beta_1, \beta_2, \beta_3 \cdots \beta_k$ are constants determined by the manufacturing process. The effective refractive index $n_{\mathrm{e}'}(r, \xi)$ experienced by the $e$ ray at the local coordinates $(r, \theta, \varsigma)$ inside the GRIN lens is represented by:

$$n_{\mathrm{e}'}(r, \xi) = \frac{n_{\mathrm{e}}(r) n_{\mathrm{o}}(r)}{\sqrt{n_{\mathrm{e}}^2(r) \cos^2 \xi + n_{\mathrm{o}}^2(r) \sin^2 \xi}}, \quad (11)$$

where $\xi$ is the interior angle between the wave normal and the extraordinary axis (blue double-sided arrow in Supplementary Figure 3b (iv)), which is the complementary angle to $\varsigma$. Along the sinusoidal ray trace, there will be an accumulated phase difference $\sigma$ between the $o$ rays and $e$ rays when the beam reaches the back surface of the GRIN lens. So, there would be a different overall $\sigma$ dependent on the traced ray $C$.

It should be noted here that there would also exist a minor beam split between $o$ rays and $e$ rays due to the birefringence properties of the GRIN lens, with which we are able to modulate the axial resolution of the GRIN lens imaging system as validated in Fig. 3b and 3c in the main article (the point spread function measurement process is adopted with reference to Ref [15]). This splitting is due to the differing sinusoidal optical path lengths of the rays ($C_e$ and $C_o$ respectively). The parameter $A$ in Eq. (9), which is related to the amplitude (and period) of the sinusoidal ray trace, is also different for $o$ rays and $e$ rays (in practice, the parameter should be represented by $A_o$ and $A_e$, separately, which are determined by the corresponding refractive index profile, Eq. (10)). As the collection of $o$ rays and $e$ rays inside GRIN lens are associated with radial and azimuthal linear polarisation eigenmodes, the corresponding linear polarized light fields play a special role in all GRIN lens based systems. Based upon this understanding, we aim to calculate the optical path length difference (*OPLD*) and the phase difference determined by the retardance σ:

$$OPLD(r,\theta,\varsigma) = \int_{C_e}[n_{e'}(s)]ds - \int_{C_o}[n_o(s)]ds \approx \int_C [n_{e'}(s) - n_o(s)]ds. \quad (12)$$

$$\sigma(r,\theta,\varsigma) = \frac{2\pi}{\lambda} \cdot \int_C [n_{e'}(s) - n_o(s)]ds. \quad (13)$$

Here $n_{e'}(s)$ and $n_o(s)$ are the local refractive indices experienced by $e$ rays and $o$ rays, as a function of distance $s$ along the sinusoidal optical path, as an arc length integral from the original point on the front surface to the exit point on the back surface. Since the birefringence in the GRIN lens is very small, when calculating the accumulated retardance, we can make the approximation that the path of the $e$ rays and the $o$ rays is the same (as shown in Eq. (12)). We use parallel incident light ($\varsigma = 0$ at z = 0) throughout this work, and as the refractive index and birefringence profiles are rotationally symmetric, if we define the wavelength of the incident beam as $\lambda$, Eq. (13) would allow us to calculate the corresponding $\sigma$.

There are several points that should be considered when estimating the uncertainties of the beam split for different GRIN systems. In this validation work, based upon our polarization measurements above, we estimated that the GRIN lenses we used had maximum birefringence level around $10^{-5}$, which is in the range of plausible values for the lithium ion-exchange process used in their manufacture. We tested three different types of pitch 2 GRIN lenses, which, due to their manufacturing processes, had different birefringence properties and lengths. We found that they exhibited variations in beam splitting ranging between ~1.5µm and ~3.0µm. These variations are due to the different optical properties of each lens. We have assumed here that $n_o(r)$ and $n_e(r)$ follow a parabolic approximation, which one could likely refine for more accurate modelling of the phenomena.

## Supplementary Note 3: Experimental set-up for polarization field measurements

Supplementary Figure 4 demonstrates the set-up for polarization field measurements. Here we used a LED light source (633nm, 3mW, $\Delta\lambda$=20nm), a polarizer (P1) and a quarter waveplate (QWP1) to generate a light field of uniform polarization that was incident on the GRIN lens cascade. We used a quarter waveplate (QWP2) and a polarizer (P2) to measure the Stokes vectors of the light field by rotating QWP2 to four different angles[17-19], following the process according to Ref [17-19, 36]. The principal equations for calculation of the Stokes vector light field are shown in Eq. (14) and Eq. (15).

$$S_{\mathrm{out}}^{n} = M_{\mathrm{P2}} M_{\mathrm{QWP2}}^{n} S_{\mathrm{in}}, \ (n = 1,2,3,4 \ldots) \tag{14}$$

$$I = A \cdot S_{\mathrm{in}}, \qquad S_{\mathrm{in}} = A^{-1} \cdot I. \tag{15}$$

where $S_{\mathrm{in}}$ is the Stokes vector of the incident light field, $M_{\mathrm{P2}}$ and $M_{\mathrm{QWP2}}^{n}$ are MMs of the corresponding polarizer and waveplate. $S_{\mathrm{out}}^{n}$ is the corresponding output Stokes vector for the $n^{\mathrm{th}}$ fast axis orientation state of the quarter waveplate. $M_{\mathrm{QWP2}}^{n}$ is the MM of the quarter waveplate for the $n^{\mathrm{th}}$ fast axis orientation. $A$ is a $n\times4$ matrix known as the instrument matrix[19], which is derived from $M_{\mathrm{P2}} \cdot M_{\mathrm{QWP2}}^{n}$. $I$ is the intensity information recorded by the camera.

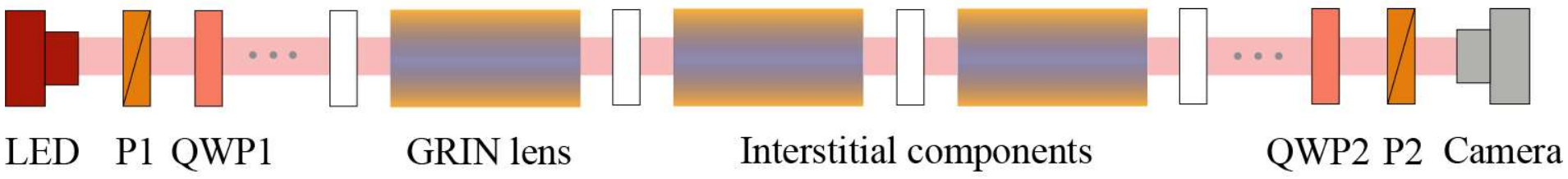

**Supplementary Figure 4. Set-up for the polarization field measurement.** LED: source; P1, P2: polarizer; QWP1, QWP2: quarter waveplate; Camera: detector. The rest of the structure is the GRIN lens cascade, which combined different kinds of GRIN lenses with interstitial optical elements. Note that imaging optics have been omitted for clarity.

## Supplementary Note 4: Theoretical and experimental validation for polarization light field characterization

Vector beams (VB) have attracted great interest for a range of applications that take advantage of their structured polarization[20-22]. They can be used to control the properties of a beam focus in microscopy[23], for the demonstration of Möbius band-like topologies[24], for electron acceleration[25] and material processing[26]. Of particular interest are singular VBs where the polarization distribution in the beam cross section has a vectorial singularity, such as a C-point (where the orientation of the polarization ellipse is undefined) or an L-line (where its handedness is indeterminate)[27, 28]. There is one class of singular VBs called the full Poincaré beam, in which the polarization structure spans the full Poincaré sphere, which are of great interest in many areas of research[27-29]. Here we first characterize the VB components of different vector vortex beams generated by GRIN lens cascade. Following the characterization of the GRIN lens polarization properties[30] using the MM in Supplementary Note 1, the properties can equivalently be represented by the Jones matrix as:

$$J_{\mathrm{GRIN}} = \begin{bmatrix} \sin^2(\theta) + \epsilon\cos^2(\theta) & (\epsilon - 1)\sin(\theta)\cos(\theta) \\ (\epsilon - 1)\sin(\theta)\cos(\theta) & \cos^2(\theta) + \epsilon\sin^2(\theta) \end{bmatrix}, \tag{16}$$

where $\epsilon = \mathrm{e}^{\mathrm{i}\sigma} = \cos(\sigma) + \mathrm{i}\sin(\sigma)$, and where the desired retardance profile $\sigma = f(r) \propto n_{\mathrm{e}}(r)$ (refer to Supplementary Note 2 for information about $f(r)$ and $n_{\mathrm{e}}(r)$ of the GRIN lens). Here, $\theta$ is the slow axis direction of the equivalent waveplate, which is equal to the angle in the cylindrical-coordinate system, $r$ is the radius, $\sigma$ is the effective cumulative retardance at position $(r,\ \theta)$. If a state of polarization (SOP) incident on the GRIN lens is described by Jones vector $E_{\mathrm{in}} = [\cos\phi, \mathrm{e}^{\mathrm{i}\delta}\sin\phi]^{\mathrm{T}}$, where the parameters $\phi$ and $\delta$ determine the polarization state, then the output polarization beam profile $E_{\mathrm{out}}$ is given by $E_{\mathrm{out}} = J_{\mathrm{GRIN}} \cdot E_{\mathrm{in}}$. This can be also represented by Eq. (17):

$$\begin{bmatrix} f_1(\phi,\delta,\theta,\sigma) \\ f_2(\phi,\delta,\theta,\sigma) \end{bmatrix} = \begin{bmatrix} \sin^2(\theta) + \epsilon\cos^2(\theta) & (\epsilon - 1)\sin(\theta)\cos(\theta) \\ (\epsilon - 1)\sin(\theta)\cos(\theta) & \cos^2(\theta) + \epsilon\sin^2(\theta) \end{bmatrix} \cdot \begin{bmatrix} \cos\phi \\ \mathrm{e}^{\mathrm{i}\delta}\sin\phi \end{bmatrix}, \tag{17}$$

where functions $f_1(\phi,\delta,\theta,\sigma)$ and $f_2(\phi,\delta,\theta,\sigma)$ determine the output VB. As for the case in Fig 2a in the main article, when we apply Eq. (17) to a right hand circular polarized incident beam, $E_{\mathrm{R}} = \frac{1}{\sqrt{2}}[1, \mathrm{i}]^{\mathrm{T}}$, it can be easily validated that when $\sigma = \pi$, the polarization states of the generated beam vary from right-hand circular (at the centre) to left-hand circular (at the outermost ring). Thus, the polarization state of the field $[\,f_1(\phi,\delta,\theta,\sigma)\,,f_2(\phi,\delta,\theta,\sigma)\,]^{\mathrm{T}}$ gradually varies across the transverse plane from the centre to the outermost boundary such that the field contains all polarization states. This validates that the single GRIN lens cascade can produce a full Poincaré beam for any pure input SOP. Eq. (17) can be similarly used to simulate the effects of other input polarization states. We demonstrated this ability more by using the same single GRIN lens cascade (see Fig. 2a in the main article as well as Supplementary Figure 5a (i) below). Supplementary Figure 5a (ii) shows polarization singularities in the exemplar full Poincaré beam (with topological index[31] $\eta$=1), that features C-point and an L-line. The corresponding mapping is shown in Supplementary Figure 5a (iii). A quantitative comparison example for demonstration is presented in Supplementary Figure 5a (iv). The simulation and experimental data are from the beam in Supplementary Figure 5a (ii). A sample distribution of polarizations along the L-line[27, 28] is chosen in order to illustrate the quality of fit; the demonstration on the Poincaré sphere is also given. The mean measurement errors of parameters *S'1, S'2* and *S'3* were 1.8%, 2.1% and 2.7% respectively (normalized by *S'0*). It should be pointed out that the precision benefits from the highly-symmetric birefringence, which is a consequence of the highly symmetric refractive index generated in the manufacturing process. That deserves further exploration, especially if seeking to use GRIN optics as a beam generator for applications that require highly symmetric beams. More examples of beam generation under this cascade are shown in Supplementary Figure 5b through simulation and experiment, which show close correspondence.

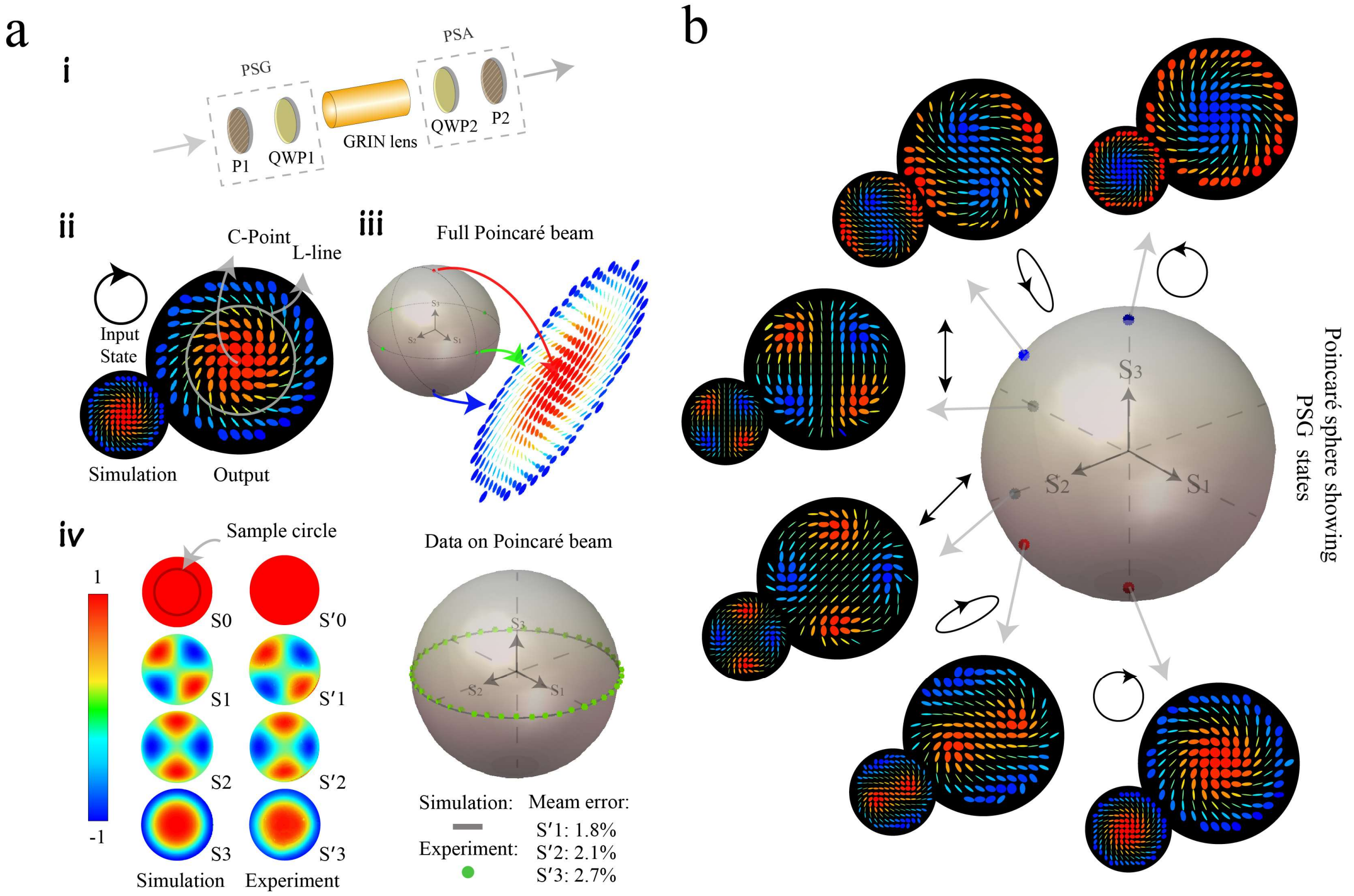

**Supplementary Figure 5. Polarization light field generation and characterization.** (**a**) (i) Schematic of a single GRIN lens cascade. P1 and P2: fixed polarizers with transmission axis at 0°. QWP1 and QWP2: rotated quarter wave plate. Before the GRIN lens is a PSG that includes P1 and QWP1 to generate a chosen arbitrary polarization state. Following the GRIN lens is a PSA, including P2 and QWP2 that enables Stokes vector measurement. (ii) shows a full Poincaré beam generated by right hand circular input, and its C-point and L-line. The projection of this beam onto Poincaré sphere are demonstrated in (iii). (iv) Stokes vector distribution of the vector beam in (ii) with both simulated and experimental data, alongside the data distribution on the Poincaré sphere. A sample distribution of polarizations along the L-line is chosen for quantitative comparison. The mean measurement errors of parameters *S'1*, *S'2* and *S'3* are given (normalized by *S'0*). (**b**) Beam generated by a single GRIN lens cascade. Five input SOPs generated by the PSG are illustrated (including linear, circular and elliptical states) and shown on the Poincaré sphere. Patterns of polarized light fields in the larger circles represent experimental results and the smaller circles are simulation counterparts. The red side of the color scale indicates right hand circular, whereas blue corresponds to left hand circular.

We further demonstrated and analysed some representative higher-order VBs generated by different cascades – with different combinations of waveplates or different GRIN lenses. An infinite range of cascades can be created, not only through different sequences of GRIN lenses and waveplates, but also through variation of the retardance and fast axis directions of the waveplates. The main principle of VB generation can be found in Eq. (18), where $E_{in}$ is the incident polarization light field and $E_{\text{out}}$ is the generated beam profile, both represented by Jones vectors. $J_{\text{inter}}^{m}$ and $J_{\text{GRIN}}^{m}$ $(m = 1,2,3 \dots)$ are $m^{\text{th}}$ Jones matrices of the interstitial optical elements and the GRIN lenses along the cascade. The simulation approach in this section is established through the methods in Supplementary Note 2 with the combination of Eq. (17) and (18).

$$E_{\text{out}} = J_{\text{inter}}^{m} \cdot J_{\text{GRIN}}^{m} \cdot J_{\text{inter}}^{m-1} \cdots J_{\text{inter}}^{2} \cdot J_{\text{GRIN}}^{1} \cdot J_{\text{inter}}^{1} \cdot E_{\text{in}}. \qquad (m = 1,2,3 \ldots) \tag{18}$$

Supplementary Figure 6 shows how different GRIN cascades can be used to generate a range of full Poincaré beams. Supplementary Figure 6a illustrates a type of GRIN lens (coloured red) that has a larger magnitude retardance profile and is almost equivalent to a sequence of three of the first type GRIN lens (shown in yellow, as also used in main article Fig. 2). Supplementary Figure 6b shows the generated beams using right hand circular polarization incident light.

Based on these results in Supplementary Figure 6b, we can make several observations:

1) The GRIN lens cascades are able to generate high radial order VBs due to the nonlinear variation of retardance along the radial direction (Supplementary Figure 6b (ii)), and to create high azimuthal order beams (Supplementary Figure 6b (i)).
2) These beams in Supplementary Figure 6b contains different kinds of topological structures[32-34], including the so-called lemon and star (associated with topological indices $\pm\frac{1}{2}$) as well as spiral (associated with topological index 1), and so on.
3) Supplementary Figures 6b (i) to (iii) show GRIN lens cascade generated multiple full Poincaré beams. In Supplementary Figure 6b (i), there are four sectors in the beam profile, each of which contains a full Poincaré beam (the white and green dotted regions show the boundaries of some of these sectors). In Supplementary Figure 6b (ii), the red dotted circle contains a full Poincaré beam, the beam inside blue dotted circle contains all polarization states spanned on the Poincaré sphere twice. In Supplementary Figure 6b (iii), we find that in a single-beam there are pair of full Poincaré beams, each with an opposite topological charge (the white and green dotted lines show one pair) and opposite handedness on the corresponding topological charge units.
4) The complex full Poincaré beam in Supplementary Figure 6b (iv) contains multiple examples of the same/opposite topological charge units[32-34] and multiple same/opposite handedness beam units (white ellipses demarcate beams that have the same handedness but with opposite topologic charges in solid and dotted outlines; those with green outlines have opposite handedness to those in white outlines but with opposite topologic charge pairs in their own solid and dotted outlines).

To our knowledge, it has not yet been reported that GRIN lenses can act as a beam generator for various types of VBs and especially for full Poincaré beams. All of the above characteristics indicate that the various special beams generated by GRIN lens cascades may open new windows for complex polarization coding and beam engineering.

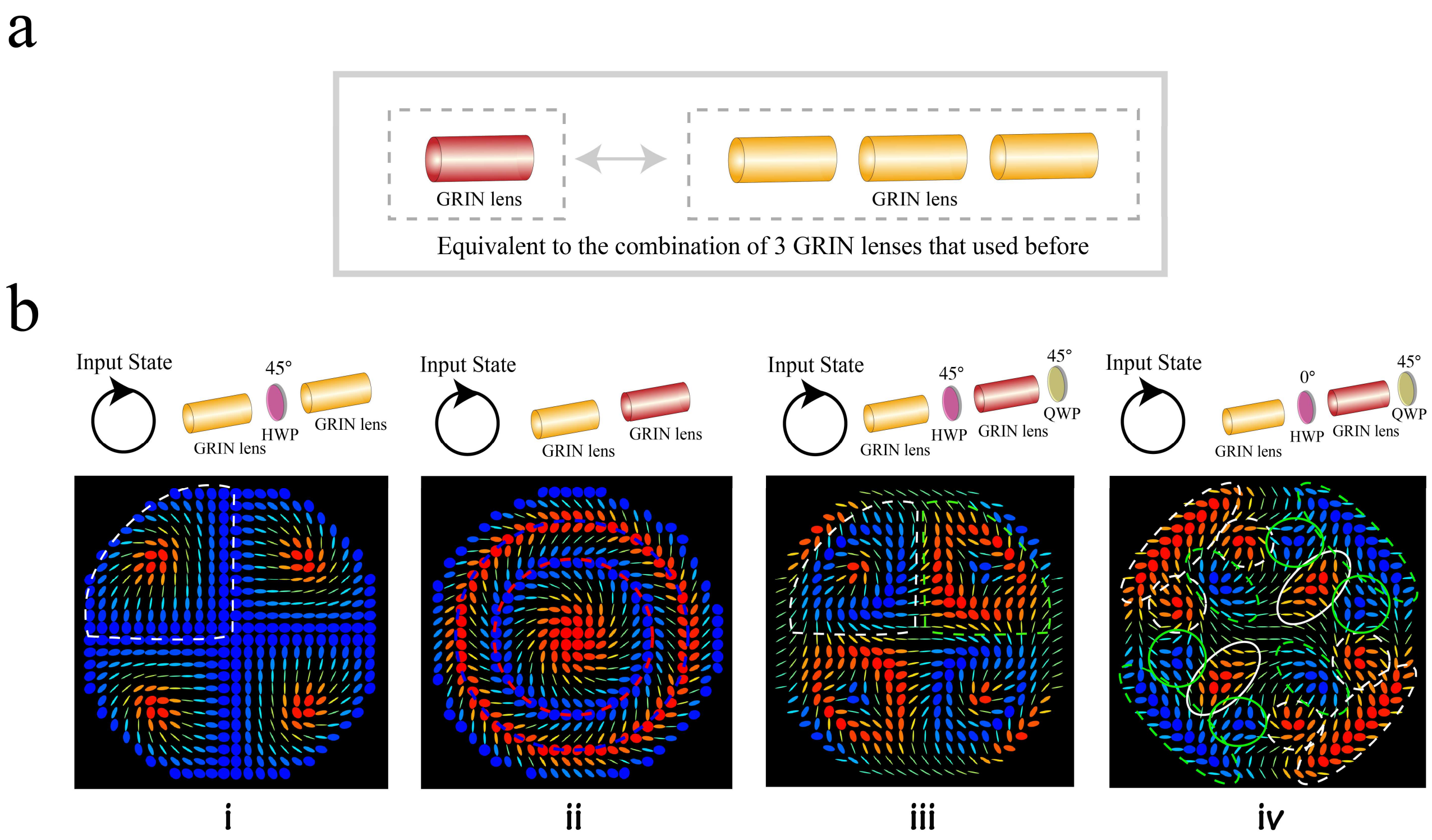

**Supplementary Figure 6. Polarization patterns generated by GRIN cascades employing two identical/different GRIN lenses.** (**a**) A special GRIN lens used in the later cases that is equivalent to an assembly of three of the GRIN lenses used in the experiment for (**b**). (**b**) are composed of four different cascade structures (from (i) to (iv)) which use identical lenses (yellow one) or different lenses (red one). These cascades were also used in Fig. 2 in the main article. These beams were generated using right hand circular polarized incident light.

## Supplementary Note 5: Set-up for interferometry

We employed commonly used interferometric methods[35, 36] to characterize OAM generation. Supplementary Figure 7 illustrates the Mach-Zehnder interferometer, in which we used a He-Ne laser (633nm, 2mW, with a Gaussian intensity distribution), a polarizer (P1) and a quarter waveplate (QWP1) to generate a uniform polarization incident light field at the input to the GRIN lens cascade. The light reflected by the first beam splitter (BS1) into the reference arm, passed off a silver mirror (M1) then was modulated by a half waveplate (HWP1). The beam was expanded then reflected again by another silver mirror (M2) before being combined with the beam from the experimental arm by the second beam splitter (BS2). Finally, the second quarter waveplate (QWP2) and the polarizer (P2) filtered the polarization state of the beam before measurement of the interferogram at the camera.

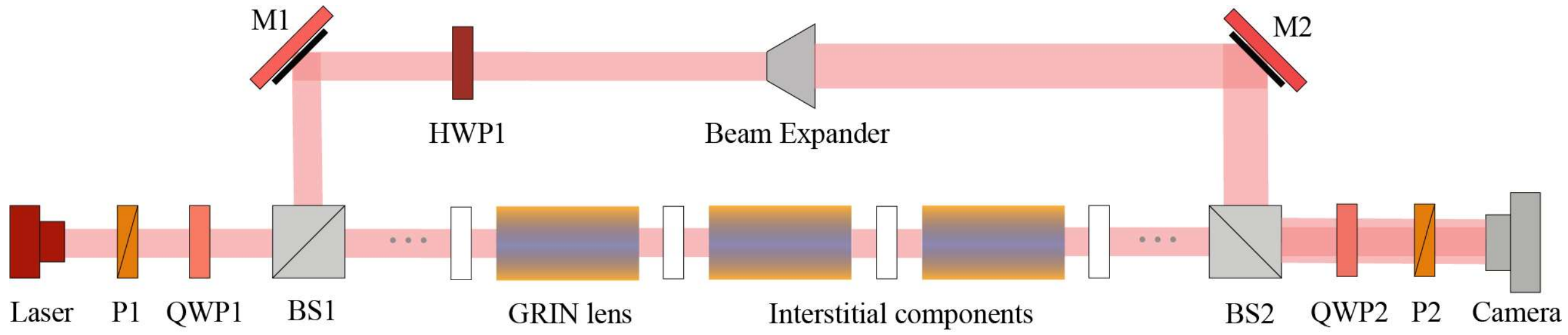

**Supplementary Figure 7. Set-up for interferometry.** P1, P2: polarizer; QWP1, QWP2: quarter waveplate; HWP1: half waveplate; BS1, BS2: beam splitter; M1, M2: silver mirror; Camera: detector. The rest of the structures comprise the GRIN lens cascade. Note that imaging optics have been omitted for clarity.

## Supplementary Note 6: Theoretical and experimental basis of OAM component analysis

A light beam possessing a helicoidal phase-front, described by the structure of $e^{il\vartheta}$, carries OAM of $l\hbar$ (where $l$ is an integer and ħ is the reduced Planck's constant)[37-40]. Since this revelation in 1992[37], various further applications of these beams have been developed, such as optical manipulation[41-44], optical communications[45-47], and in quantum and nano-optics[48-51]. In this section, based on the polarization pattern analysis in above section, we provide further details of the analysis of OAM components of different vector vortex beams generated by GRIN lens cascade. From Eq. (16) we know that the Jones Matrix of a GRIN lens is equal to:

$$J_{\mathrm{GRIN}} = \frac{1}{2}\begin{bmatrix} J_{11} & J_{12} \\ J_{21} & J_{22} \end{bmatrix} = \begin{bmatrix} \sin^2(\theta) + \epsilon\cos^2(\theta) & (\epsilon-1)\sin(\theta)\cos(\theta) \\ (\epsilon-1)\sin(\theta)\cos(\theta) & \cos^2(\theta) + \epsilon\sin^2(\theta) \end{bmatrix}, \tag{19}$$

where $\epsilon = \mathrm{e}^{\mathrm{i}\sigma} = \cos(\sigma) + \mathrm{i}\sin(\sigma)$, and $\sigma = f(r) \propto n_{\mathrm{e}}(r)$ (see Supplementary Note 2), $\theta$ is the fast axis direction of the equivalent waveplate which equals to the azimuthal angle in the cylindrical coordinate system, $r$ is the radius across the section of the GRIN lens, $\sigma$ is the linear retardance value of the equivalent waveplate at the position $(r, \theta)$. For simplicity of notation, let us make $S^{2\theta} = \sin(2\theta)$, $C^{2\theta} = \cos(2\theta)$, $S^{\sigma} = \sin(\sigma)$, $C^{\sigma} = \cos(\sigma)$. From this we obtain:

$$\begin{bmatrix} J_{11} & J_{12} \\ J_{21} & J_{22} \end{bmatrix} = \begin{bmatrix} 1 + C^{\sigma} + C^{2\theta}(C^{\sigma}-1) + S^{\sigma}(1 + C^{2\theta})\mathrm{i} & S^{2\theta}(C^{\sigma}-1) + S^{\sigma}S^{2\theta}\mathrm{i} \\ S^{2\theta}(C^{\sigma}-1) + S^{\sigma}S^{2\theta}\mathrm{i} & 1 + C^{\sigma} + C^{2\theta}(1-C^{\sigma}) + S^{\sigma}(1 - C^{2\theta})\mathrm{i} \end{bmatrix}. \tag{20}$$

Suppose a uniformly polarized beam represented by Jones vector $E_{\mathrm{in}}$ passes through the GRIN lens $J_{\mathrm{GRIN}}$. We now examine the properties of the generated vector $E_{\mathrm{out}} = J_{\mathrm{GRIN}} \cdot E_{\mathrm{in}}$ in left- and right-circular eigenpolarization bases: $E_{\mathrm{L}} = \frac{1}{\sqrt{2}}[1, -\mathrm{i}]^{\mathrm{T}}$, $E_{\mathrm{R}} = \frac{1}{\sqrt{2}}[1, \mathrm{i}]^{\mathrm{T}}$. Following the OAM generation case in the main article, we used a fixed input SOP: $E_{\mathrm{R}} = \frac{1}{\sqrt{2}}[1, \mathrm{i}]^{\mathrm{T}}$ entering a single GRIN lens. Using the Jones matrix calculation of Eq. (20), we find that,

$$J_{\mathrm{GRIN}}\begin{bmatrix} 1 \\ \mathrm{i} \end{bmatrix} = z_1\begin{bmatrix} 1 \\ \mathrm{i} \end{bmatrix} + z_2\begin{bmatrix} 1 \\ -\mathrm{i} \end{bmatrix}, \tag{21}$$

where $z_1$ and $z_2$ can be written as $z_1 = \frac{1}{2}(\mathrm{e}^{\mathrm{i}0} + \mathrm{e}^{\mathrm{i}\sigma})$, $z_2 = \frac{1}{2}\left(\mathrm{e}^{\mathrm{i}(2\theta+\pi)} + \mathrm{e}^{\mathrm{i}(2\theta+\sigma)}\right)$. So, if we use a PSA for state $E_{\mathrm{L}} =$ $\frac{1}{\sqrt{2}}[1, -\mathrm{i}]^{\mathrm{T}}$, would only see the $z_2 E_{\mathrm{L}}$ term from Eq. (21), which can be written as,

$$z_2 \begin{bmatrix} 1 \\ -\mathrm{i} \end{bmatrix} = \frac{1}{2}\left(\mathrm{e}^{\mathrm{i}(2\theta+\pi)} + \mathrm{e}^{\mathrm{i}(2\theta+\sigma)}\right) \begin{bmatrix} 1 \\ -\mathrm{i} \end{bmatrix} = A\, \mathrm{e}^{\mathrm{i}(2\theta+\varphi)} \begin{bmatrix} 1 \\ -\mathrm{i} \end{bmatrix}, \tag{22}$$

where $A(r)$ is the amplitude distribution. When considering OAM, we are only concerned about the phase term $(2\theta + \varphi)$ in Eq. (22), where $\varphi(r)$ is an initial phase delay determined by the retardance profile $\sigma(r)$ (see Supplementary Note 2). The exponent $\mathrm{i}(2\theta)$ shows that the analyzed $E_{\mathrm{L}}$ beam exhibits two units of OAM, which reveals that the GRIN lens can be used as a spin-to-orbital angular momentum convertor. Corresponding results can be found in Fig. 2a (iii) to (v), Supplementary Figures 8a and 8c.

We now examine the properties of the generated vector $E_{\mathrm{out}} = J_{\mathrm{GRIN}} \cdot E_{\mathrm{in}}$ in another two specific eigenpolarization bases: $E_{\mathrm{H}} = [1,0]^{\mathrm{T}}$, $E_{\mathrm{V}} = [0,1]^{\mathrm{T}}$. We used a fixed input SOP: $E_{\mathrm{R}} = \frac{1}{\sqrt{2}}[1, \mathrm{i}]^{\mathrm{T}}$ entering a single GRIN lens:

$$J_{\mathrm{GRIN}} \begin{bmatrix} 1 \\ \mathrm{i} \end{bmatrix} = z_{1'} \begin{bmatrix} 1 \\ 0 \end{bmatrix} + z_{2'} \begin{bmatrix} 0 \\ 1 \end{bmatrix}. \tag{23}$$

From calculation, $[z_{1'} \quad z_{2'}]$ can be written as exponential form:

$$z_{1'} = \frac{1}{2}\left(\mathrm{e}^{\mathrm{i}0} + \mathrm{e}^{\mathrm{i}\sigma} + \mathrm{e}^{\mathrm{i}(2\theta+\pi)} + \mathrm{e}^{\mathrm{i}(\sigma+2\theta)}\right), \tag{24}$$

$$z_{2'} = \frac{1}{2}\left(\mathrm{e}^{\mathrm{i}\left(\frac{\pi}{2}\right)} + \mathrm{e}^{\mathrm{i}\left(\sigma+\frac{\pi}{2}\right)} + \mathrm{e}^{\mathrm{i}\left(2\theta+\frac{\pi}{2}\right)} + \mathrm{e}^{\mathrm{i}\left(\sigma+2\theta-\frac{\pi}{2}\right)}\right). \tag{25}$$

In Supplementary Figure 8 below, the corresponding phase profiles, intensity profiles, as well as interference patterns using analyzers described by $[1, 0]^{\mathrm{T}}$ (both in experiment and simulation) can be found in Supplementary Figures 8d (iii) to 8h (iii) and 8d (vii) to 8h (vii). We also modified the PSA to obtain a wider variety of wave fronts for illustration; the sequence is shown in Supplementary Figure 8b on the pathway from (i) to (ix) on the Poincaré sphere of the PSA state. The simulated and experimental interferograms as well as intensity distributions can be found in Supplementary Figures 8d to 8h.

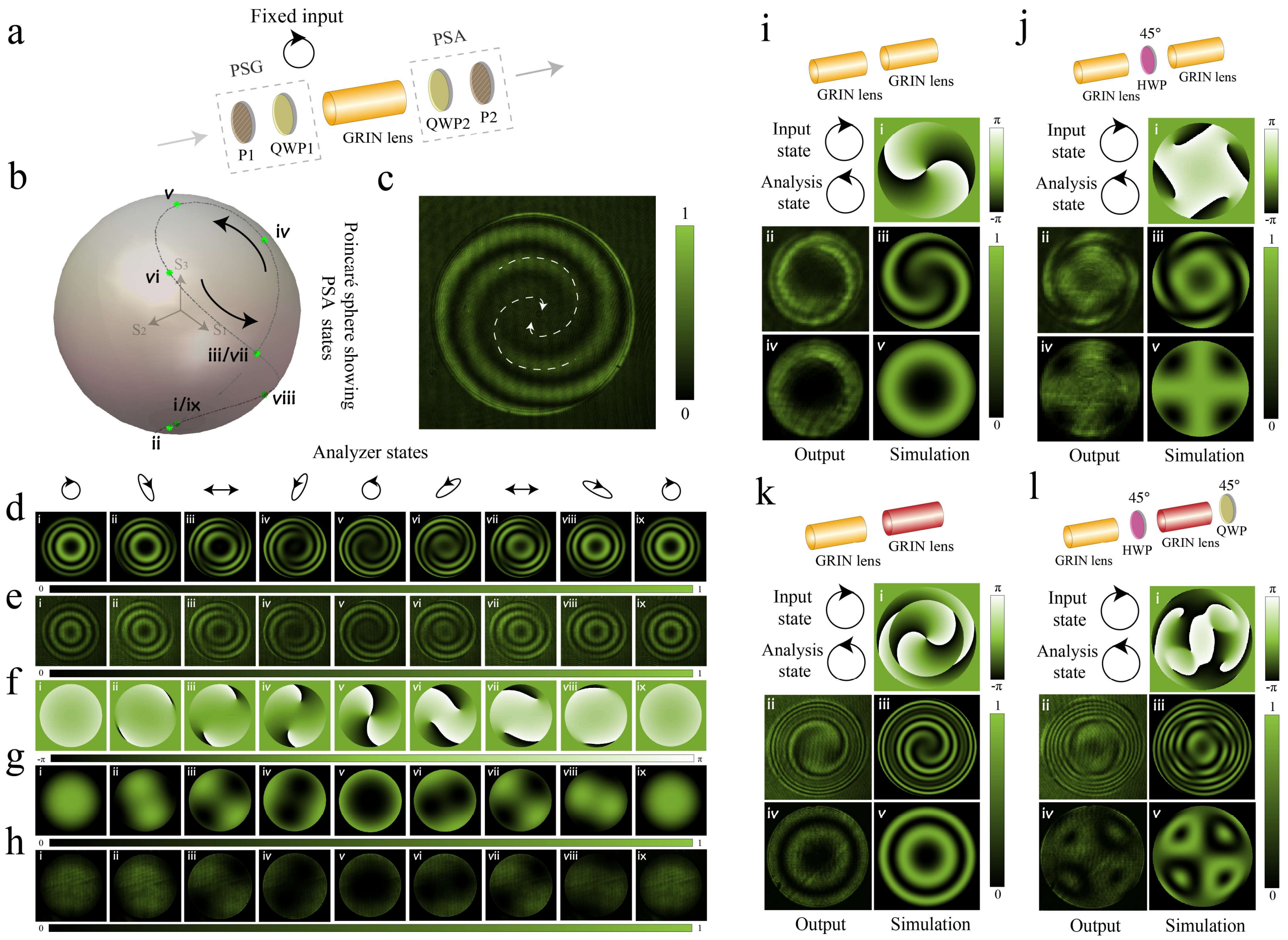

**Supplementary Figure 8. OAM generation and characterization.** (**a**) Simplified setup for the OAM generation validation. P1 and P2: fixed polarizers with transmission axis at 0°. QWP1: fixed quarter wave plate with fast axis orientation at 45°. QWP2: rotating quarter wave plate. The PSG comprises P1 and QWP1 to generate fix circular SOP, and the PSA is constructed by P2 and QWP2. (**b**) A Poincaré sphere representation of the pathway of detection polarization states (i) to (ix) in (**d**) to (**h**). (**c**) A particular illustration of the interference pattern of state (*v*) in (**f**). The two spirals indicated that the light beam contained two units of OAM. (**d**) to (**h**) Interference patterns, phase profiles, intensity distributions under different fast axis orientations of the second quarter waveplate (QWP2 in (**a**)) from -45° to 135° with interval 22.5°. The rows show: (**d**) Simulation of interference patterns. (**e**) Experimental interference patterns. (**f**) Simulation of phase patterns. (**g**) Simulation of intensity distribution. (**h**) Experimental intensity distribution. The analyser states for each column are shown above (**b**). The data here are normalized besides the simulation phase value. (**i**), (**j**), (**k**) and (**l**) are higher-order GRIN lens cascades and their generated complex phase patterns, interference patterns and intensity profiles under specific input SOP and analysis polarization state. Note that the PSG and PSA parts have been omitted for clarity, and that in (**i**) and (**j**) the second GRIN lens (yellow) is of the same type as the first lens, whereas in (**k**) and (**l**) GRIN lens (red) has a larger magnitude retardance profile, as used in the previous section.

Similar to the polarization fields shown in the main article (Fig. 2), some demonstrations of the complex phase modulation abilities of higher order GRIN lens cascades are also shown in Supplementary Figures 8i to 8l. It can be seen that higher order cascades are able to generate different complex phase patterns that also feature higher order OAM components by choosing the input SOP and the analysis polarization state. An example can been seen in Supplementary Figure 8j, which shows four discrete singularities. This method can benefit various applications requiring phase control, especially for OAM or vortex beam related applications.

We further demonstrated cases using linear polarized incident eigenpolarization bases: $E_{\mathrm{H}} = [1{,}0]^{\mathrm{T}}$, $E_{\mathrm{V}} = [0{,}1]^{\mathrm{T}}$. We first used a fixed input SOP: $E_{\mathrm{H}} = [1{,}0]^{\mathrm{T}}$ entering a single GRIN lens cascade, and analyzed by $E_{\mathrm{H}} = [1{,}0]^{\mathrm{T}}$ and $E_{\mathrm{V}} = [0{,}1]^{\mathrm{T}}$.

$$J_{\mathrm{GRIN}}\begin{bmatrix}1\\0\end{bmatrix} = z_{1''}\begin{bmatrix}1\\0\end{bmatrix} + z_{2''}\begin{bmatrix}0\\1\end{bmatrix}. \tag{26}$$

We calculate $z_{1''}$ and $z_{2''}$ by through the same process used before, as shown in Eq. (27) and (28). In the following Supplementary Figures 9a and 9b we demonstrate phase profiles, intensity distributions and interference patterns (where the object beam is interfered with the same SOP light).

$$z_{1''} = \begin{cases} \frac{1}{2}\left(\mathrm{e}^{\mathrm{i}0} + \mathrm{e}^{\mathrm{i}\sigma}\right) + \frac{1}{2}|\cos(2\theta)|\left(\mathrm{e}^{\mathrm{i}0} + \mathrm{e}^{\mathrm{i}(\sigma+\pi)}\right), & \theta \in \left[-\frac{3\pi}{4}, -\frac{\pi}{4}\right] \cup \left[\frac{\pi}{4}, \frac{3\pi}{4}\right], \\ \frac{1}{2}\left(\mathrm{e}^{\mathrm{i}0} + \mathrm{e}^{\mathrm{i}\sigma}\right) + \frac{1}{2}|\cos(2\theta)|\left(\mathrm{e}^{\mathrm{i}\pi} + \mathrm{e}^{\mathrm{i}\sigma}\right), & \text{Otherwise.} \end{cases} \tag{27}$$

$$z_{2''} = \begin{cases} \frac{1}{2}|\sin(2\theta)|\left(\mathrm{e}^{\mathrm{i}0} + \mathrm{e}^{\mathrm{i}(\sigma+\pi)}\right), & \theta \in \left[-\frac{\pi}{2}, 0\right] \cup \left[\frac{\pi}{2}, \pi\right], \\ \frac{1}{2}|\sin(2\theta)|\left(\mathrm{e}^{\mathrm{i}\pi} + \mathrm{e}^{\mathrm{i}\sigma}\right), & \text{Otherwise.} \end{cases} \tag{28}$$

Using the basis $E_{\mathrm{R}} = \frac{1}{\sqrt{2}}[1, \mathrm{i}]^{\mathrm{T}}$, $E_{\mathrm{L}} = \frac{1}{\sqrt{2}}[1, -\mathrm{i}]^{\mathrm{T}}$ as the analysis eigenbasis, we obtain:

$$J_{\mathrm{GRIN}}\begin{bmatrix}1\\0\end{bmatrix} = z_{1'''}\begin{bmatrix}1\\\mathrm{i}\end{bmatrix} + z_{2'''}\begin{bmatrix}1\\-\mathrm{i}\end{bmatrix}, \tag{29}$$

from which we derive Eq. (30) and Eq. (31) below. The corresponding phase profiles, intensity distributions and interference patterns can be found in Supplementary Figures 9c and 9d.

$$z_{1'''} = \frac{1}{4}\left(\mathrm{e}^{\mathrm{i}0} + \mathrm{e}^{\mathrm{i}\sigma} + \mathrm{e}^{-\mathrm{i}(2\theta+\pi)} + \mathrm{e}^{\mathrm{i}(\sigma-2\theta)}\right), \tag{30}$$

$$z_{2'''} = \frac{1}{4}\left(\mathrm{e}^{\mathrm{i}0} + \mathrm{e}^{\mathrm{i}\sigma} + \mathrm{e}^{\mathrm{i}(2\theta+\pi)} + \mathrm{e}^{\mathrm{i}(\sigma+2\theta)}\right). \tag{31}$$

The simulations in this section were based on the above theoretical equations. It should be noted that other combinations of incident polarizations or analysis eigenvectors, or even higher order GRIN lens cascades cases (combined with Eq. (18)), can be easily calculated through the same process above.

We have demonstrated here that the GRIN lens cascades have the potential to generate complex structured phase profiles under different input/output eigenpolarizations. This novel structure for phase modulation (including the generation of OAM), derives from the special birefringence profile of GRIN optics. To our knowledge, this has not yet been reported. It might pave the way for new methods of complex beam engineering and benefit corresponding applications such as OAM related quantum and nano-optics.

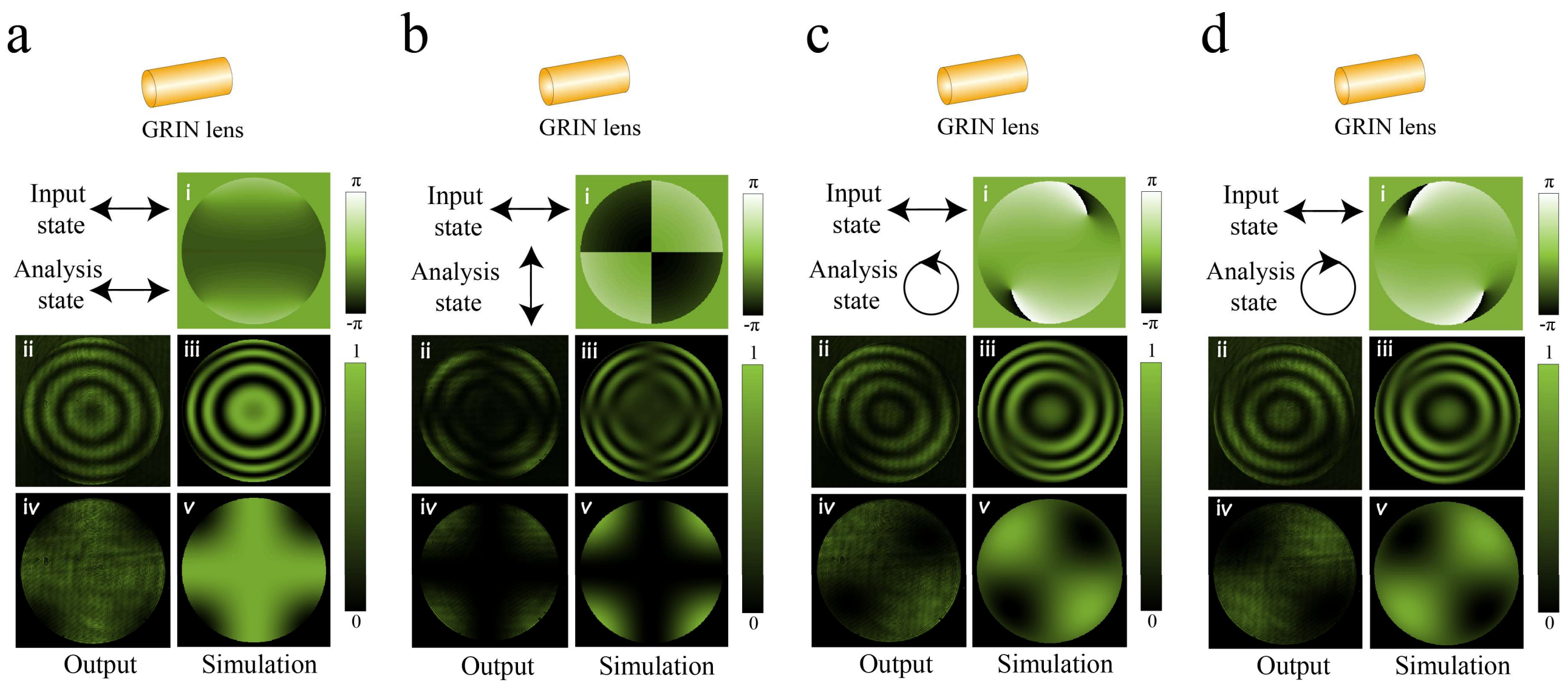

**Supplementary Figure 9. Phase profiles, interference patterns, intensity distributions generated by the single GRIN lens.** (**a**), (**b**), (**c**) and (**d**) are phase and intensity information extracted using horizontal linear polarized light at the input and analysis through: (**a**) horizontal linear, (**b**) vertical linear, (**c**) left hand circular and (**d**) right hand circular polarized light, respectively. In each group of four images, (i) shows simulated absolute phase profiles, (ii) and (iii) are experimental and simulated interference patterns, (iv) and (v) are experimental and simulated intensity distributions. For clarity, the PSG and PSA parts are not shown in the cascades.

## Supplementary Note 7: Optimization of the polarisation state generator (PSG) and polarisation state analyser (PSA)

There are many ways to specify the illumination polarizations and analyser configurations for MM polarimetry[1, 3-5, 7-11, 13, 18, 52, 53]. To find optimum measurement configurations, one can use the condition number (CN) of the system as an optimisation criterion for both the PSG and PSA. The CN can also be related to the volume of the inner tetrahedroid constructed by the SOPs inside the Poincaré sphere (shown in Fig. 4b (ii) of the main article)[18, 52].

Based upon this, we chose a PSG that consisted of four sectors of differently oriented quarter waveplates, as shown in Fig. 4b (i) in the main article, where the fast axes of waveplates 1 to 4 were oriented at 45°, 15.9°, 74.1°, and -45° according to the optimization shown in Ref [52]. Similar CN analysis has shown that the 132° phase retardance with four different fast axis directions is optimal for the PSA[18, 52], providing a CN of 1.732. As shown in Fig. 4c (ii) and 4c (iii) in the main article, the polarization properties of the GRIN lens cascade have rotational symmetry (this cascade is also used to generate multiple full Poincaré beams units in a single beam in Supplementary Note 4). However, each of the four sectors equivalently encompasses required optimal combinations of retardance and fast axis direction, and each sector is enabled to generate a full Poincaré beam. Additionally, as the birefringence profile is determined by the atomic level scale of ion implantation, the number of potential channels in the system is determined in effect by the pixelation of the camera. Thus, the GRIN lens based MM polarimeter has the great advantage of providing multiple parallel channels to aid the calculation. We chose $m$ ($m$=4, 5, 6…) pixels in each channel to calculate four Stokes vectors simultaneously, to then work out the target MM. If taking MMPD results as a standard, the chosen pixels encompass the analysis channels of the 132° phase retardance ring on each PSA sector, so it follows that the optimum CN (1.732) can be achieved. The additional effective pixels can help reduce spatial noise when combining the pixel information[18]. For these reasons, the GRIN cascade can form as a basis of an optimal MM measurement system.

## Supplementary Note 8: Measurement of the Müller matrix (MM) with the single shot polarimeter

We derive here the main equations describing the principle of operation of the MM polarimeter. We use *n* (*n*=1, 2, 3 and 4) to denote the four areas of the PSG array corresponding to the four sectors of the FQWP, and *m* (*m*=1, 2, 3 ... *M*) to denote the chosen pixel number in each sector.

$$S_{\mathrm{out}}^{n,m} = M_{\mathrm{P2}} \cdot M_{\mathrm{GRIN2}}^{n,m} \cdot M_{\mathrm{HWP}} \cdot M_{\mathrm{GRIN1}}^{n,m} \cdot M_{\mathrm{Sample}} \cdot S_{\mathrm{in}}^{n,m}, \tag{32}$$

where $S_{\mathrm{in}}^{n,m}$ represents incident Stokes vectors generated by the four sectors of the PSG array (the *m* is the chosen pixel number in each sector; all pixels in each sector have same incident SOP). $S_{\mathrm{out}}^{n,m}$ is the combination of output Stokes vectors (both in the channel *m* of the sector *n*; the same meanings are followed in the later description). $M_{\mathrm{Sample}}$ denotes the MM of the targeted sample, and $M_{\mathrm{P2}}$, $M_{\mathrm{HWP}}$ denote the MMs of the polarizer and the half wave plate. $M_{\mathrm{GRIN1}}^{n,m}$, $M_{\mathrm{GRIN2}}^{n,m}$ are MMs of the GRIN lenses in the corresponding spatial positions. Since only intensity information can be recorded by the camera, which means only the first element of $S_{\mathrm{out}}^{n,m}$ can be obtained (the intensity information at sector *n* from channel *m*), then Eq. (32) can be modified into Eq. (33) and Eq. (34):

$$I_{\mathrm{out}} = A \cdot M_{\mathrm{Sample}} \cdot S_{\mathrm{in}}, \tag{33}$$

$$\begin{bmatrix} I_{\mathrm{out}}^{1,1} \\ \vdots \\ I_{\mathrm{out}}^{1,M} \\ I_{\mathrm{out}}^{2,1} \\ \vdots \\ I_{\mathrm{out}}^{2,M} \\ I_{\mathrm{out}}^{3,1} \\ \vdots \\ I_{\mathrm{out}}^{3,M} \\ I_{\mathrm{out}}^{4,1} \\ \vdots \\ I_{\mathrm{out}}^{4,M} \end{bmatrix} = \begin{bmatrix} a_0^{1,1} & a_1^{1,1} & a_2^{1,1} & a_3^{1,1} \\ & & \vdots & \\ a_0^{1,M} & a_1^{1,M} & a_2^{1,M} & a_3^{1,M} \\ a_0^{2,1} & a_1^{2,1} & a_2^{2,1} & a_3^{2,1} \\ & & \vdots & \\ a_0^{2,M} & a_1^{2,M} & a_2^{2,M} & a_3^{2,M} \\ a_0^{3,1} & a_1^{3,1} & a_2^{3,1} & a_3^{3,1} \\ & & \vdots & \\ a_0^{3,M} & a_1^{3,M} & a_2^{3,M} & a_3^{3,M} \\ a_0^{4,1} & a_1^{4,1} & a_2^{4,1} & a_3^{4,1} \\ & & \vdots & \\ a_0^{4,M} & a_1^{4,M} & a_2^{4,M} & a_3^{4,M} \end{bmatrix} \cdot M_{\mathrm{Sample}} \cdot \begin{bmatrix} s_0^{1,1} & \cdots & s_0^{1,M} & s_0^{2,1} & \cdots & s_0^{2,M} & s_0^{3,1} & \cdots & s_0^{3,M} & s_0^{4,1} & \cdots & s_0^{4,M} \\ \vdots & \ddots & \vdots & \vdots & \ddots & \vdots & \vdots & \ddots & \vdots & \vdots & \ddots & \vdots \\ s_3^{1,1} & \cdots & s_3^{1,M} & s_3^{2,1} & \cdots & s_3^{2,M} & s_3^{3,1} & \cdots & s_3^{3,M} & s_3^{4,1} & \cdots & s_3^{4,M} \end{bmatrix}. \tag{34}$$

Where the intensity information recorded by the camera for each modulation and detection channel are defined as a vector:

$$I_{\mathrm{out}} = \begin{bmatrix} I_{\mathrm{out}}^{1,1} & \cdots & I_{\mathrm{out}}^{1,M} & I_{\mathrm{out}}^{2,1} & \cdots & I_{\mathrm{out}}^{2,M} & I_{\mathrm{out}}^{3,1} & \cdots & I_{\mathrm{out}}^{3,M} & I_{\mathrm{out}}^{4,1} & \cdots & I_{\mathrm{out}}^{4,M} \end{bmatrix}^{\mathrm{T}}, \tag{35}$$

and $A$ is defined by:

$$A = [A^{1,1} \quad \cdots \quad A^{1,M} \quad A^{2,1} \quad \cdots \quad A^{2,M} \quad A^{3,1} \quad \cdots \quad A^{3,M} \quad A^{4,1} \quad \cdots \quad A^{4,M}]^{\mathrm{T}}, \tag{36}$$

which is a $4m \times 4$ matrix in which each component $A^{n,m} = [a_0^{n,m} \quad a_1^{n,m} \quad a_2^{n,m} \quad a_3^{n,m}]$ consists of the first row of the corresponding $M_{\mathrm{P2}} \cdot M_{\mathrm{GRIN2}}^{n,m} \cdot M_{\mathrm{HWP}} \cdot M_{\mathrm{GRIN1}}^{n,m}$. We then expand $S_{\mathrm{in}}^{n,m} = [s_0^{n,m} \quad s_1^{n,m} \quad s_2^{n,m} \quad s_3^{n,m}]^{\mathrm{T}}$ , and let

$$S_{\mathrm{in}} = \begin{bmatrix} s_0^{1,1} & \cdots & s_0^{1,M} & s_0^{2,1} & \cdots & s_0^{2,M} & s_0^{3,1} & \cdots & s_0^{3,M} & s_0^{4,1} & \cdots & s_0^{4,M} \\ \vdots & \ddots & \vdots & \vdots & \ddots & \vdots & \vdots & \ddots & \vdots & \vdots & \ddots & \vdots \\ s_3^{1,1} & \cdots & s_3^{1,M} & s_3^{2,1} & \cdots & s_3^{2,M} & s_3^{3,1} & \cdots & s_3^{3,M} & s_3^{4,1} & \cdots & s_3^{4,M} \end{bmatrix}, \tag{37}$$

which is a $4 \times 4m$ matrix that consists of columns that are individual Stokes vectors from each combination of PSG sectors and camera pixels. Then the MM of the sample can be calculated as:

$$M_{\mathrm{Sample}} = A^{-1} \cdot I_{\mathrm{out}} \cdot S_{\mathrm{in}}^{-1}, \tag{38}$$

where $S_{\mathrm{in}}^{-1}$ is the pseudo inverse matrix of $S_{\mathrm{in}}$ , and $A^{-1}$ is the pseudo inverse matrix of $A$.

## Supplementary Note 9: Validation of feasibility using standard samples

In order to test the capabilities of the MM polarimeter, we used as samples four polarizers with different orientations (0°, 90°, 45°, -45°), which are illustrated as P2 to P5 in the experimental setup in Supplementary Figure 10a. We took measurements of each polarizer in a single-shot and compared the derived MM elements with the ground truth MMs (Supplementary Figures 10b (i) to 10b (iv)). The maximum errors of the derived MM elements were smaller than 6.93%. The GRIN lens cascade based polarimeter has the following advantages: 1) it can perform single-shot MM measurement in a robust and stable manner; 2) single-shot measurements enable precise measurement of moving/changing objects; 3) it has the potential to be miniaturised into an integrated instrument, especially a fibre-based probe with scanning detection for clinical applications.

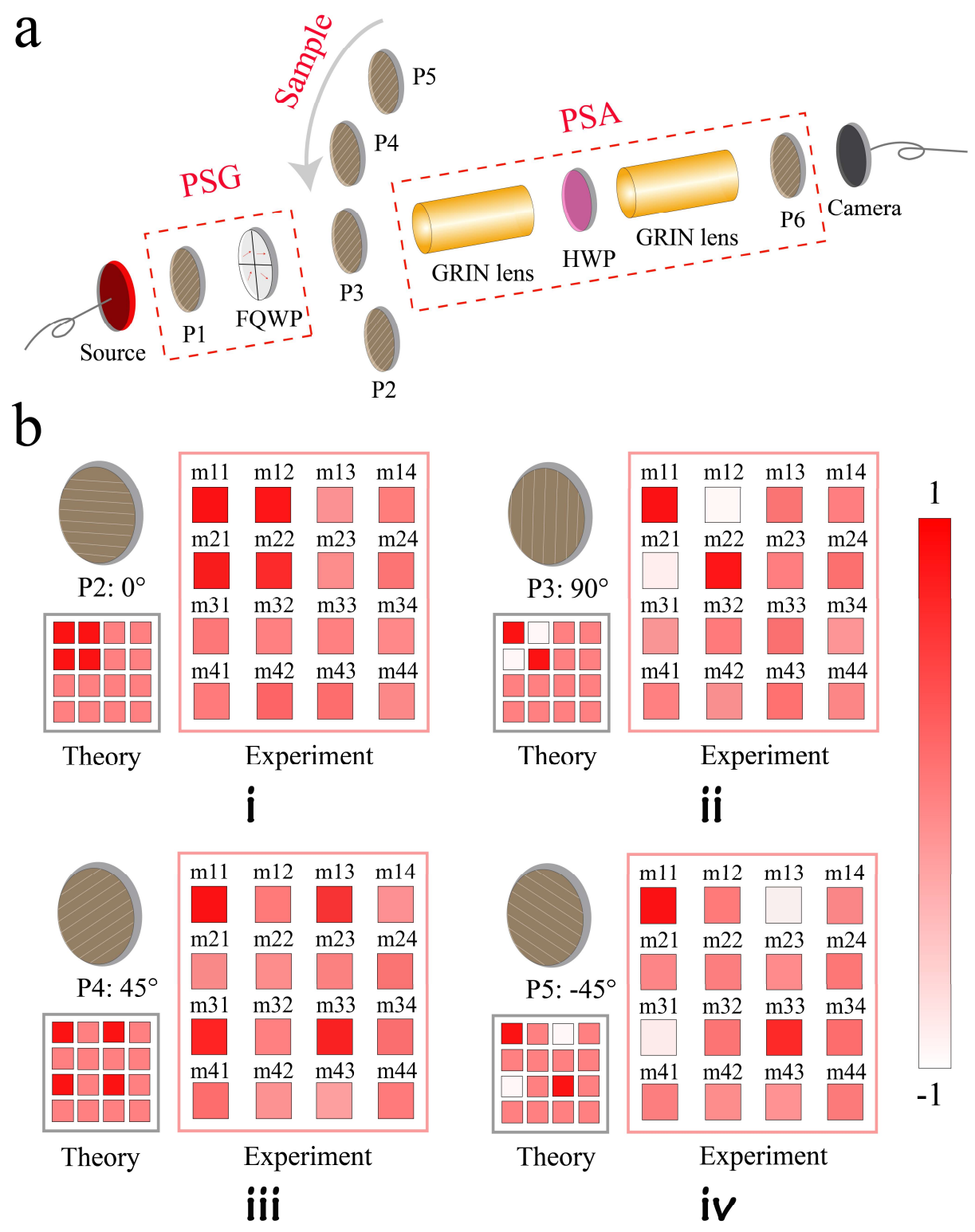

**Supplementary Figure 10. Characterization of the GRIN lens based MM polarimeter.** (**a**) Simplified setup for the validation experiment of the polarimeter. PSG: P1, polarizer; FQWP, four-quadrant quarter waveplate, with different fast axis orientations in each quadrant. PSA: P6, polarizer; HWP, half waveplate. P2 to P5: polarizers with transmission axis orientations at 0°, 90°, 45°, and -45°. (**b**) (i) to (iv) The measured single-shot MMs of the four polarizers (red boxes), and corresponding ground truth (black boxes).

## Supplementary Note 10: Tissue sample information, sampling method and statistical analysis for tissue measurements

Clinically, different stages of breast ductal carcinoma tissues have different proportions and distribution of fibrous structures in and around the milk ducts (there are stages 1 (normal), 2 (ductal carcinoma in situ) and 3 (invasive ductal carcinoma))[53]. Through this work, the samples were unstained, dewaxed sections of human breast ductal carcinoma tissue slices at different stages. They were prepared and provided by the Department of Pathology, Shenzhen Sixth People's (Nanshan) Hospital. For the tissues in stages 1 (normal), and 3 (invasive ductal carcinoma), we selected 12-μm-thick slices[53] for demonstration. For comparison, the corresponding 4-μm-thick hematoxylin and eosin (H and E) stained slices[53] were also prepared. The age range of the patients was from thirty to fifty-five years. This work was approved by the Ethics Committee of the Shenzhen Sixth People's (Nanshan) Hospital.

We measured 10 points per sample using our new MM polarimeter and a conventional MM microscope (overall 20 samples), as ground truth for quantitative comparison. In each sample, we chose the points by using MM polarimeter with the diameter of the field of view (FOV) at 0.19mm (Femto Technology Co. Ltd., G-B151157-S1483). The FOV of the MM microscope (calibration precision $< 0.3\%$) was around four times larger than that of the polarimeter, so we took measurements from chosen sub-areas within same FOV of MM polarimeter to be processed. Then we calculated the mean value of the retardance across these areas to set as ground truth for further comparison through the same quantitative comparison process used to differentiate the breast cancerous stages by polarization parameters[53]. Example MMs, as well as corresponding MMPD parameters, from healthy or cancerous tissue are illustrated alongside quantitative statistic histograms in Supplementary Figure 11.

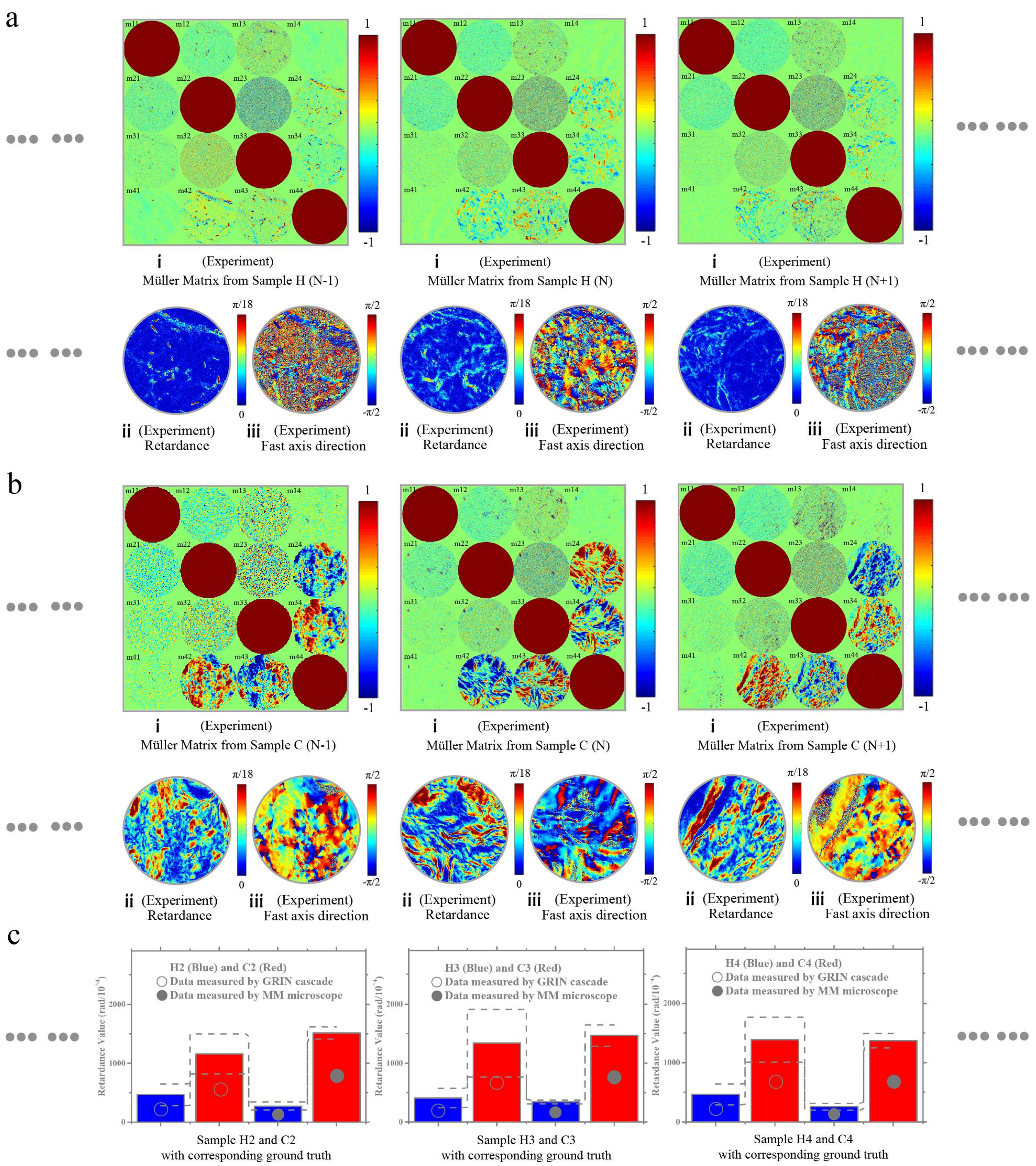

**Supplementary Figure 11. Original MMs of healthy and cancerous samples.** (**a**) Demonstrations of the original MMs (i), MMPD parameters (retardance (ii) and fast axis directions (iii)) of the healthy samples. (**b**) Demonstrations of the original MMs, MMPD parameters (retardance (ii) and fast axis directions (iii)) of the cancerous samples. (**c**) Demonstrations of statistic histograms (mean value and the standard deviation of the retardance) of different samples measured by the GRIN lens cascade and a conventional MM microscope (ground truth); numbers refer to Supplementary Table 1. Source data are provided as a Source Data file.

For simplicity of measurement by the MM polarimeter we make the assumption that measurement of the average SOP should be sufficient for discrimination purposes. We first defocused the GRIN lens image, in order to implement optical averaging of the polarization effects of spatially varying structures. Then we began with a chosen region with optimal CN (related to the 132° retardance[18]) to calculate the MM for demonstration of feasibility. If strong non-uniform area occurred such as near to the boundary between tissue types, we chose alternative rings (corresponding to other retardance values within the effective CN region on the PSA[52]) or applied another single-shot to reconstruct the MM. In principle, alternative region shapes can be chosen for optimization of the processes.

All the obtained data measured by the MM polarimeter were used to draw an overall 3-D dot distribution in Fig. 4f in the main article. Then, combined with the corresponding MM microscope data, Supplementary Table 1 was formed to show the data measured by two approaches as well as the corresponding P-value[53], which shows the significant difference between the two classes of samples. Considering the projection on Y (sample number) - Z (retardance) plane in the main article (Fig. 4f) and the P-values in Supplementary Table 1 of either individual samples or the overall combination, it can be seen that the polarimeter is able to distinguish efficiently the healthy and cancerous tissues. The difference between two kinds of samples is obvious across the tested samples; this also confirms the robustness of the polarimeter. Compared to the data measured by the MM microscope, larger standard deviations from the data measured by the polarimeter can be observed. The difference might result from the inhomogeneous sample structure in the effective FOV. To make the process more efficient and precise, further detailed error analysis and corresponding optimization will be the subject of further work.

**Supplementary Table 1. Value of retardance δ (rad) of samples used in this section**

| | Slide H1 and Slide C1 | | | | | Slide H2 and Slide C2 | | | | |
|---|---|---|---|---|---|---|---|---|---|---|
| | *Healthy* | | *Cancerous* | | | *Healthy* | | *Cancerous* | | |
| | *mean value* | *standard deviation* | *mean value* | *standard deviation* | **P-value* | *mean value* | *standard deviation* | *mean value* | *standard deviation* | **P-value* |
| **GRIN cascade** | 0.0365 | 0.0166 | 0.1412 | 0.0497 | <0.001 | 0.0463 | 0.0185 | 0.1159 | 0.0338 | <0.001 |
| **MM microscope** | 0.0309 | 0.0089 | 0.1289 | 0.0151 | <0.001 | 0.0271 | 0.0067 | 0.1513 | 0.0104 | <0.001 |
| | **Slide H3 and Slide C3** | | | | | **Slide H4 and Slide C4** | | | | |
| | *Healthy* | | *Cancerous* | | | *Healthy* | | *Cancerous* | | |
| | *mean value* | *standard deviation* | *mean value* | *standard deviation* | **P-value* | *mean value* | *standard deviation* | *mean value* | *standard deviation* | **P-value* |
| **GRIN cascade** | 0.0406 | 0.0164 | 0.1340 | 0.0573 | <0.001 | 0.0464 | 0.0178 | 0.1385 | 0.0377 | <0.001 |
| **MM microscope** | 0.0314 | 0.0033 | 0.1467 | 0.0179 | <0.001 | 0.0257 | 0.0056 | 0.1371 | 0.0122 | <0.001 |
| | **...** | | | | | **Combination of all slides** | | | | |
| | *Healthy* | | *Cancerous* | | | *Healthy* | | *Cancerous* | | |
| | *mean value* | *standard deviation* | *mean value* | *standard deviation* | **P-value* | *mean value* | *standard deviation* | *mean value* | *standard deviation* | **P-value* |
| **GRIN cascade** | **...** | | | | | 0.0408 | 0.0189 | 0.1324 | 0.0433 | <0.001 |
| **MM microscope** | | | | | | 0.0323 | 0.0102 | 0.1279 | 0.0171 | <0.001 |

*the P-value is obtained by the significance test method[53] ($P<=0.05$ is significant), which shows the significance of the difference between the two sets of data.

## Supplementary Reference